# Τα δίκτυα μεταφορών στην Ελλάδα και η σημασία τους για την οικονομική ανάπτυξη

Δημήτριος Τσιώτας, Μάρθα Γεράκη, Σπύρος Νιαβής

Τμήμα Μηχανικών Χωροταξίας Πολεοδομίας και Περιφερειακής Ανάπτυξης,
Πανεπιστήμιο Θεσσαλίας, Πεδίον Άρεως, Βόλος, 38 334,
Τηλ +30 24210 74446, fax: +302421074493
E-mails: tsiotas@uth.gr; mageraki@uth.gr

**Περίληψη**
Το άρθρο αυτό επιχειρεί να αναδείξει τη σημασία που κατέχουν οι μεταφορές στην οικονομική ανάπτυξη της Ελλάδας και ειδικότερα οι μεταφορικές υποδομές και τα δίκτυα, τα οποία συνιστούν ένα πάγιο δομημένο κεφάλαιο που εκτείνεται, με διάφορες μορφές, στο σύνολο της χώρας. Για το σκοπό αυτό, μελετώνται διαχρονικά και διατομεακά στατιστικά στοιχεία που περιγράφουν ορισμένα θεμελιώδη μακροοικονομικά μεγέθη και μέτρα. Περαιτέρω, η μελέτη επιχειρεί να επισημάνει τη δομική και τη λειτουργική διάσταση που συνθέτουν την έννοια των δικτύων μεταφορών και να αναδείξει την αναγκαιότητα της από κοινού θεώρησής τους στην επιστημονική έρευνα. Τα δίκτυα μεταφορών που εξετάζονται είναι το οδικό, το σιδηροδρομικό, το ακτοπλοϊκό και το αεροπορικό δίκτυο της Ελλάδας, τα οποία προσεγγίζονται τόσο ως προς τη γεωμετρία και τα τεχνικά χαρακτηριστικά τους, όσο και ως προς το ιστορικό, κυκλοφοριακό και πολιτικό τους πλαίσιο. Για την εμπειρική αποτίμηση της σημασίας των μεταφορικών δικτύων στην Ελλάδα κατασκευάζεται ένα οικονομετρικό υπόδειγμα που εκφράζει το επίπεδο ευημερίας των νομών ως συνάρτηση των μεταφορικών τους υποδομών και του κοινωνικοοικονομικού τους πλαισίου. Απώτερο σκοπό του άρθρου αποτελεί η ανάδειξη, συνδυαστικά, του συνόλου των πτυχών που συνοδεύουν τη μελέτη των μεταφορικών υποδομών και δικτύων.

**Λέξεις Κλειδιά:** οδικό δίκτυο, σιδηροδρομικό δίκτυο, ακτοπλοϊκό δίκτυο, αεροπορικό δίκτυο.

**Abstract**
This article attempts to highlight the importance that transportation has in the economic development of Greece and in particular the importance of the transportation infrastructure and transportation networks, which suggest a fixed structured capital covering the total of the country. For this purpose, longitudinal and cross-sectoral statistical data are examined over a set of fundamental macroeconomic measures and metrics. Furthermore, the study attempts to highlight the structural and functional aspects composing the concept of transportation networks and to highlight the necessity of their joint consideration on the relevant research. The transportation networks that are examined in this paper are the Greek road (GRN), rail (GRAN), maritime (GMN) and air transport network (GAN), which are studied both in terms of their geometry and technical characteristics, as well as of their historical, traffic and political framework. For the empirical assessment of the transportation networks' importance in Greece an econometric model is constructed, expressing the welfare level of the Greek regions as a multivariate function of their transportation infrastructure and of their socioeconomic environment. The further purpose of the article is to highlight, macroscopically, all the aspects related the study of transportation infrastructure and networks.

## 1. Εισαγωγή

Η μετακίνηση των ανθρώπων αποτελεί ιστορικό φαινόμενο, το οποίο εμφανίστηκε από την έμφυτη ανάγκη για *επικοινωνία* και *κινητικότητα* και το οποίο επιδρά διαχρονικά στην οικονομική και κοινωνική ανάπτυξη των κοινωνιών (Πολύζος, 2011). Εντός αυτού του εννοιολογικού πλαισίου, οι μεταφορές συνιστούν το σύνολο των οργανωμένων ανθρώπινων δράσεων, οι οποίες αποσκοπούν στην ικανοποίηση της ανάγκης του ανθρώπου για επικοινωνία και κινητικότητα και αφορούν την κάλυψη των χωρικών αποστάσεων και γενικά των χωρικών περιορισμών που εμφανίζονται από τη διασπορά των θέσεων των κοινωνικών σχηματισμών και των κοινωνικών δραστηριοτήτων.

Οι μεταφορές, οι οποίες υπήρξαν ανέκαθεν αποτύπωμα των ανθρώπινων δραστηριοτήτων στο χώρο, είναι δυνατόν να διακριθούν σε τρεις βασικές κατηγορίες (Πολύζος, 2002, 2003, 2005, 2011):

- *Μεταφορά ανθρώπων*: Η κατηγορία αυτή αφορά την ανάγκη μετακίνησης των ατόμων μεταξύ περιοχών για διάφορους σκοπούς.
- *Μεταφορά αγαθών*: Αυτή η κατηγορία αναφέρεται στη μετακίνηση των υλικών αγαθών, με σκοπό την κάλυψη των γεωγραφικών αποστάσεων για την προώθηση της παραγωγικής διαδικασίας και την πραγματοποίηση των εμπορικών συναλλαγών.
- *Μεταφορά πληροφοριών, γνώσης και τεχνολογίας*: Η κατηγορία αυτή αφορά τη διάδοση των πληροφοριών και της τεχνολογικής γνώσης μεταξύ διαφορετικών χωρικών μονάδων, ως αποτέλεσμα της αλληλεπίδρασης της κουλτούρας και του πολιτισμού που συντελείται κατά την επικοινωνία.

Ιστορικά, παρατηρείται ότι η ανάπτυξη των μεταφορών και των μεταφορικών υποδομών είναι στενά συνδεδεμένη με την ανάπτυξη των ανθρώπινων κοινωνιών, αποτελώντας το αποτύπωμά τους στο χώρο. Διαχρονικά έχει αποδειχθεί ότι η *χωρική αντίσταση* (Tsiotas and Polyzos, 2013b), στην οποία υπόκεινται οι μεταφορές, διαρκώς μειώνεται, προάγοντας τις δυνατότητες επικοινωνίας και οικονομικών αλληλεπιδράσεων (Πολύζος, 2011). Επίσης, προκύπτει ιστορικά ότι η ανάπτυξη των μεγάλων αστικών κέντρων είναι στενά συνδεδεμένη με τη γεωγραφική τους θέση, αλλά και με τη δυνατότητα αξιοποίησης των υφιστάμενων φυσικών μεταφορικών συστημάτων (παραποτάμιες ή παραθαλάσσιες πόλεις) ή με τη δυνατότητα ανάπτυξης ανθρωπογενών μεταφορικών υποδομών (Πολύζος, 2011). Γενικά, οι υποδομές των μεταφορών αποτελούν το δομικό υπόβαθρο, εντός του οποίου υλοποιούνται οι δράσεις που συντελούνται σε κάθε μία από τις προαναφερόμενες κατηγορίες των μεταφορών και το οποίο έχει διαδραματίσει διαχρονικά σημαντικό ρόλο στην ανάπτυξη των κοινωνικών σχηματισμών, όπως είναι οι πόλεις, οι περιφέρειες ή σε ευρύτερη κλίμακα οι χώρες (Πολύζος, 2003, 2011).

Η σχέση ανάμεσα στις μεταφορές και την οικονομική ανάπτυξη χαρακτηρίζεται από υψηλό βαθμό πολυπλοκότητας. Με όρους Περιφερειακής Οικονομικής (Behrens and Thisse, 2007), οι μεταφορές συντελούν διαχρονικά στη σύνδεση των χρήσεων γης, προάγοντας την παραγωγική διαδικασία, στην υλοποίηση των εμπορικών συναλλαγών με τη δημιουργία ροών εμπορίου (Πολύζος, 2005), στην ανάπτυξη των τοπικών οικονομιών, στη μείωση των διαπεριφερειακών ανισοτήτων και γενικότερα της χωρικής ασυμμετρίας (Πολύζος, 2002), στη μεγέθυνση των εθνικών οικονομιών και, ευρύτερα, στην προαγωγή των διεθνών οικονομικών συναλλαγών. Οι μεταφορές μπορεί να επηρεάσουν τη διανομή του εισοδήματος, το μείγμα της παραγωγής και των εισροών, το επίπεδο απασχόλησης και την τεχνολογική πρόοδο. Σε πολλές περιπτώσεις, η ύπαρξη μη ανεπτυγμένου μεταφορικού συστήματος ανάμεσα σε μη ανεπτυγμένες περιφέρειες αντικατοπτρίζει την ανεπάρκεια της αντίστοιχης ζήτησης (Πολύζος, 2011). Οι μεταφορές συμβάλλουν καθοριστικά στην προώθηση των οικονομικών αλληλεπιδράσεων (οικονομική ανάπτυξη), αλλά και στην προώθηση της διαπροσωπικής επικοινωνίας (κοινωνική ανάπτυξη).

Στον αντίποδα της άποψης ότι οι μεταφορικές υποδομές προάγουν ευθέως την περιφερειακή ανάπτυξη υφίσταται προσεγγίσεις που διατυπώνουν ότι οι υποδομές των μεταφορών συντελούν στην αύξηση των περιφερειακών ανισοτήτων. Ο Drew (1990) υποστηρίζει ότι το σύστημα μεταφορών δεν αποτελεί ικανή και αναγκαία συνθήκη για την ανάπτυξη των περιφερειών, αναγνωρίζει όμως ότι βοηθά στην παράκαμψη των χωρικών εμποδίων και δημιουργεί προϋποθέσεις περαιτέρω ανάπτυξης. Ως διαδικασία, η οικονομική μεγέθυνση έπειτα από την κατασκευή μεταφορικών υποδομών εμφανίζεται αρχικά στις αστικοποιημένες ή πλησίον αστικών περιοχών περιφέρειες, διότι διαθέτουν τους απαραίτητους ανθρώπινους και φυσικούς πόρους και την απαιτούμενη οικονομική δομή για την άμεση εκμετάλλευση των υποδομών αυτών. Μέσα από μια αλυσιδωτή πορεία, οι επιχειρηματίες επηρεάζονται από την πλευρά τους στο να εγκαταστήσουν τις επιχειρήσεις τους στις περιφέρειες που αναπτύσσουν συγκριτικό πλεονέκτημα ως προς το μεταφορικό κόστος, λόγω της προσβασιμότητας στις αγορές και τις πηγές προμήθειας πρώτων υλών. Ωστόσο, τα συνολικά αποκομισμένα οφέλη ποικίλουν κατά περίπτωση βιομηχανίας λόγω της διαφορετικής εξάρτησής τους από το μεταφορικό κόστος. Καθώς οι μεταφορικές υποδομές συνεισφέρουν στην αύξηση της παραγωγικότητας, τότε προάγουν και την ανάπτυξη. Ο Plassard (1992) θεωρεί ότι η ύπαρξη θετικής σχέσης μεταξύ των μεταφορικών υποδομών και της περιφερειακής ανάπτυξης στερείται επαρκούς επιστημονικής τεκμηρίωσης και ότι οι απόψεις που την υποστηρίζουν συντηρούνται λόγω της ανάγκης νομιμοποίησης πολιτικών αποφάσεων με σκοπό την κατασκευή μεταφορικών υποδομών και την ύπαρξη συνέχειας μεταξύ βραχυχρόνιων και μακροχρόνιων μεταβολών, οι οποίες ήταν δύσκολο να διακριθούν μεταξύ τους. Εκτιμά, επίσης, ότι η κατασκευή μεταφορικών υποδομών δεν δημιουργεί αξιοσημείωτες μακροχρόνιες περιφερειακές μεταβολές.

Με μία συνθετική προσέγγιση, ο Rephann (1993) διακρίνει 3 ανταγωνιστικές απόψεις σχετικά με τις βέλτιστες περιφερειακές συνθήκες στις οποίες θα πρέπει να υλοποιούνται μεταφορικές υποδομές, ώστε να μεγιστοποιείται η αποτελεσματικότητά τους. Η πρώτη άποψη υποστηρίζει ότι οι υποδομές αυτές πρέπει να κατασκευάζονται στις αναπτυσσόμενες περιφέρειες, η δεύτερη προτείνει ενδιάμεσες περιφέρειες και η Τρίτη βασίζεται στη θεωρία των πόλων ανάπτυξης, σύμφωνα με την οποία η κατασκευή των μεταφορικών υποδομών πρέπει να επικεντρώνεται σε περιοχές κυρίως με αστική σύνθεση πληθυσμού, οι οποίες εμφανίζουν προηγούμενο αναπτυξιακό δυναμισμό.

Ο Vickerman (1989, 1995) θεωρεί ότι πρέπει να εξετάζονται σε βάθος περαιτέρω παράγοντες από την τυπική προσέγγιση που επικεντρώνεται στην προσβασιμότητα και στο αποτέλεσμα που επιφέρει η μεταβολή του μεταφορικού κόστους στις βιομηχανίες της περιφέρειας. Κατά το συγγραφέα, πρέπει να εξετάζεται το ενδεχόμενο δημιουργίας «αποτελεσμάτων διαδρόμου» (corridor effects), ώστε μια περιφέρεια να μην επηρεάζεται από την διέλευση που προκαλείται μέσω αυτής της μεταφορικής υποδομής ώστε να περιορίζεται να εκμεταλλευθεί τα πλεονεκτήματα που προσφέρει η υποδομή. Χαρακτηριστικά παραδείγματα κατά το Vickerman αποτελούν οι περιοχές "Kent", της Μ. Βρετανίας και "Nord Pas de Calais", της Γαλλίας, οι οποίες βρίσκονται στις δύο εισόδους του Channel Tunnel. Ο συγγραφέας επίσης αναγνωρίζει ως πρόβλημα την υπόθεση της στατικής οικονομικής δομής των περιφερειών.

Στον ελληνικό χώρο, ο Πετράκος (1997) κωδικοποιώντας τις παραπάνω προσεγγίσεις αναφέρει ότι τα μεταφορικά δίκτυα μπορούν να συντελέσουν στην περιφερειακή ανάπτυξη σε συνδυασμό με τους άλλους τρεις αναπτυξιακούς παράγοντες μιας περιοχής, οι οποίοι είναι η θέση της στο γεωγραφικό χώρο, οι υφιστάμενες οικονομίες συγκέντρωσης και η τομεακή σύνθεση της τοπικής παραγωγής και απασχόλησης. Το μέγεθος των αποκομισμένων ωφελειών μιας μη κεντρικής περιφέρειας που μειώνει την απόσταση της λόγω ανάπτυξης μεταφορικών υποδομών, εξαρτάται κυρίως από την παραγωγική της βάση και από την ικανότητα αντίστασή της προς την ελκτική δύναμη που θα δεχθεί από τις κεντρικές για απορρόφηση της αγοραστικής της δύναμης και του εργατικού της δυναμικού και συνεπώς αρκετών οικονομικών της δραστηριοτήτων. Οι οι επιδράσεις μπορούν να είναι θετικές ή αρνητικές και η γεωγραφική θέση (κεντρική ή περιμετρική) μαζί με τον δυναμισμό της οικονομίας κάθε επηρεαζόμενης από τις διαπεριφερειακές μεταφορικές υποδομές περιοχής αποτελούν τα βασικά χαρακτηριστικά γνωρίσματα που καθορίζουν την κατεύθυνση αυτή των επιπτώσεων.

Με όρους της Επιστήμης των Δικτύων (Network Science) (Lewis, 2011; Brandes et al., 2013), οι μεταφορές συνιστούν συστήματα αλληλεπίδρασης και επικοινωνίας, των οποίων οι υποδομές αναπαριστούν δίκτυα και οι μετακινήσεις εκφράζουν τις ροές που συντελούνται εντός των δικτύων. Για παράδειγμα, στις οδικές μεταφορές οι οδικές υποδομές συνιστούν ένα δίκτυο, εντός του οποίου αναπτύσσονται ροές μετακίνησης που εξυπηρετούν κάθε μία από τις προαναφερόμενες κατηγορίες των μεταφορών. Όμοια, το σύστημα ναυτιλιακών μεταφορών που συνίσταται από το σύνολο των ενεργών λιμένων μιας χώρας και των ακτοπλοϊκών αξόνων της, περιγράφει ένα δίκτυο των ακτοπλοϊκών μεταφορών (Tsiotas and Polyzos, 2015a), στο οποίο διεξάγονται εξίσου μεταφορικές δράσεις (ροές) και από τις τρεις κατηγορίες μεταφορών. Αντίστοιχα, το σύστημα των αεροπορικών μεταφορών, το οποίο αποτελείται από τους ενεργούς αερολιμένες μιας χώρας και από τους αεροδιαδρόμους (εναέριες τροχιές μετακίνησης), συνιστά ένα δίκτυο αεροπορικών μεταφορών (Tsiotas and Polyzos, 2015b), στο οποίο εξυπηρετούνται επίσης όλες οι προαναφερόμενες κατηγορίες, αλλά κυρίως η μεταφορά προσώπων.

Μία χαρακτηριστική άποψη που διατυπώνεται από τους Πετράκο και Ψυχάρη (2004) υποστηρίζει ότι η έμφαση που δίδεται στις μεταφορικές υποδομές είναι υπερβολική για το μερίδιο που αυτές κατέχουν στο σύνολο των συστημάτων επικοινωνίας (πχ. τηλεπικοινωνίες, ευρυζωνικά δίκτυα) και οφείλεται στην υλική (απτή) φύση τους, σε αντίθεση με τις υποδομές που θεωρούνται «άυλες» (δηλ. στις οποίες η πληροφορία/το σήμα μεταφέρεται μέσω διαύλων που δεν έχουν δομημένη υπόσταση, πχ. στον αέρα). Η άποψη αυτή αφορά σαφώς στη «στατικότητα» των μεταφορικών υποδομών έναντι των άλλων «άυλων» συστημάτων επικοινωνίας που εμφανίζουν χαρακτηριστική δυναμική και προσαρμοστικότητα στις διαχρονικές και κάθε είδους αλλαγές. Ωστόσο, η στατικότητα που τους προσδίδει η δομημένη φύση των μεταφορικών υποδομών αναδεικνύει την εξέχουσα σημασία που κατέχει ο προγραμματισμός και ο σχεδιασμός ως εργαλεία χάραξης αναπτυξιακής πολιτικής στον τομέα των μεταφορών, καθόσον ιδιαίτερα δύσκολη και δαπανηρή η εξάλειψη οποιωνδήποτε σφαλμάτων και παραλείψεων.

Από τα παραπάνω, ως *δίκτυα μεταφορών* μπορούν να οριστούν τα συστήματα των συνδέσεων που υλοποιούνται μεταξύ διαφόρων θέσεων ή χωρικών μονάδων (όπως αστικοί σχηματισμοί, πόλεις, λιμάνια, αερολιμένες), συντελούνται μέσα σε ένα συγκεκριμένο χωρικό υποδοχέα (εδαφική επιφάνεια, θαλάσσια επιφάνεια, αέρας) και αποσκοπούν στη διενέργεια των μεταφορών και κατ' επέκταση στην περιφερειακή ανάπτυξη. Η μορφή, η δομή και οι λειτουργίες των δικτύων μεταφορών ποικίλουν ανάλογα με τις εκάστοτε ιστορικές, οικονομικές και κοινωνικές συνθήκες, με το μέσο της μεταφοράς και προφανώς με το χωρικό υποδοχέα.

Τέλος, η αλληλεπίδραση των συστημάτων χρήσεων γης και των δικτύων μεταφορών αποτελεί ένα από τα βασικότερα φαινόμενα που εμφανίζονται στις περιοχές με αστικές συγκεντρώσεις. Γενικά παρατηρείται ότι, η ανάπτυξη των δικτύων μεταφοράς αποτελεί πόλο έλξης για την εγκατάσταση κοινωνικών και οικονομικών δραστηριοτήτων στις παρακείμενες περιοχές. Η ανάπτυξη περιοχών (κόμβων) υψηλής προσβασιμότητας αυξάνει κατά κανόνα τη συνολική ζήτηση για εγκατάσταση των δραστηριοτήτων, δημιουργώντας συγκριτικό πλεονέκτημα έναντι άλλων ανταγωνιστικών περιοχών. Η ανάπτυξη αυτή αποκτά πολλαπλασιαστικό χαρακτήρα, καθόσον οι περιοχές υψηλής προσβασιμότητας ανατροφοδοτούν νέα ανάπτυξη και συγκέντρωση περαιτέρω οικονομικών δραστηριοτήτων (Πολύζος, 2011).

Στο πλαίσιο αυτό, το παρόν άρθρο επιχειρεί την επισκόπηση των βασικών θεματικών αξόνων που αναδεικνύουν τη σημασία των μεταφορών και ειδικότερα των μεταφορικών υποδομών και δικτύων στην οικονομική και περιφερειακή ανάπτυξη της Ελλάδας. Τα μεταφορικά δίκτυα συνιστούν το πάγιο δομημένο κεφάλαιο που εκτείνεται με διαφορετικές μορφές στο σύνολο της χώρας με σκοπό την αλληλεπίδραση, το εμπόριο και την επικοινωνία των συνδεδεμένων περιοχών. Για το σκοπό αυτό, μελετώνται διαχρονικά και διατομεακά στατιστικά στοιχεία, τα οποία περιγράφουν ορισμένα θεμελιώδη μακροοικονομικά μεγέθη και μέτρα που σχετίζονται με τις μεταφορές. Η μεθοδολογική προσέγγιση του άρθρου στηρίζεται στην παρουσίαση και επεξεργασία των εν λόγω μεγεθών με τεχνικές της περιγραφικής και επαγωγικής στατιστικής, καθώς και στην κατασκευή ενός οικονομετρικού υποδείγματος για την εμπειρική εξέταση της συνεισφοράς των μεταφορικών υποδομών στην περιφερειακή ανάπτυξη. Περαιτέρω, το άρθρο επιχειρεί να επισημάνει τη δομική και τη λειτουργική διάσταση που συνθέτουν την έννοια των δικτύων μεταφορών και να αναδείξει την αναγκαιότητα της από κοινού θεώρησής τους στην επιστημονική έρευνα. Η εργασία επιδιώκει να αποτελέσει άρθρο επισκόπησης που να παρέχει στον αναγνώστη τη θεματική βάση για τη μελέτη των δικτύων μεταφορών.

## 2. Οι Μεταφορές στην Ελλάδα

Οι μεταφορές στην Ελλάδα αποτελούν βασική συνιστώσα της εθνικής οικονομίας και σημαντικό παράγοντα της ανάπτυξής της. Η διαπίστωση αυτή στοιχειοθετείται είτε με προφανή τρόπο, από τη μελέτη των κρατικών δομών, είτε σε επίπεδο άσκησης της οικονομικής πολιτικής της χώρας, μέσα από την αξιολόγηση των οικονομικών μεγεθών της. Σε επίπεδο εκτελεστικής εξουσίας, είναι χαρακτηριστικό ότι, διαχρονικά, στη διάρθρωση της δομής των ελληνικών κυβερνήσεων δίδεται ιδιαίτερη έμφαση στον τομέα των μεταφορών. Σύμφωνα με την υπουργική δομή της ελληνικής κυβέρνησης του έτους 2012, προκύπτει ότι 2 από τα συνολικά 18 Υπουργεία της χώρας (Υπουργείο Ναυτιλίας και Αιγαίου, Υπουργείο Υποδομών, Μεταφορών και Δικτύων) υπήρξαν αφιερωμένα αποκλειστικά στο αντικείμενο των μεταφορών. Επιπρόσθετα, δύο ακόμη Υπουργεία (Υπουργείο Δημόσιας Τάξης και Προστασίας του Πολίτη, Υπουργείο Τουρισμού) είναι επιφορτισμένα με λειτουργίες που έχουν συνάφεια με τις μεταφορές, όπως η οδική ασφάλεια και η ασφάλεια των μεταφορών (οι οποίες αποτελούν αρμοδιότητα της Διεύθυνσης Τροχαίας), αλλά και ο τουρισμός (Πολύζος, 2011, Polyzos et al., 2013a), η ανάπτυξη του οποίου εξαρτάται άμεσα από την ποιότητα των συστημάτων μεταφορών της χώρας.

Η ύπαρξη ενός ποιοτικού συστήματος μεταφορών σε μια χώρα ή γενικότερα σε μια γεωγραφική ενότητα αποτελεί ικανή αλλά όχι αναγκαία προϋπόθεση για την οικονομική ανάπτυξή της. Αντιθέτως, η ύπαρξη ενός αποδοτικού δικτύου μεταφορών αποτελεί ικανή και αναγκαία συνθήκη για τη μεγέθυνση της οικονομίας των λιγότερο αναπτυγμένων χωρών. Για το λόγο αυτό, οι εκτέλεση έργων αναβάθμισης των συγκοινωνιακών υποδομών αποτελούν κατά κανόνα πρώτιστη προτεραιότητα των αναπτυξιακών πολιτικών που εφαρμόζονται στις αναπτυσσόμενες οικονομίες (Πολύζος, 2011). Η σπουδαιότητα των μεταφορικών υποδομών στην εθνική οικονομία οδηγεί τις χώρες στην διάθεση ετησίως ενός σημαντικού μέρους των κρατικών δαπανών για τη συντήρηση και την αναβάθμισή τους (Πολύζος, 2002, 2011). Η παραπάνω διαπίστωση εφαρμόζεται και στο επίπεδο άσκησης της οικονομικής πολιτικής της Ελλάδας. Ειδικότερα, στον Πίνακα 1 παρουσιάζεται το μερίδιο του Προϋπολογισμού Δημόσιων Επενδύσεων της χώρας, που έχουν ανατεθεί προς διαχείριση στο Υπουργείο Υποδομών, Μεταφορών και Δικτύων, για την περίοδο 2009-2016.

**Πίνακας 1**[α]

Το μερίδιο του προϋπολογισμού Δημοσίων Επενδύσεων της χώρας στο Υπουργείο Υποδομών, Μεταφορών και Δικτύων, για ενδεικτικά έτη από την περίοδο 2009-2016.

|  | **2009** Απολογισμός[β] (€) |  | **2010** Απολογισμός (€) |  |
|---|---|---|---|---|
| Προϋπολογισμός Δημοσίων Επενδύσεων | 8.288.054.218,20 |  | 7.702.013.564,49 |  |
| Μερίδιο Υπουργείου Υποδομών, Μεταφορών και Δικτύων | 674.002.640,06 | (8,13%) | 2.363.342.624,46 | (30,68%) |
|  | **2012** Απολογισμός (€) |  | **2014** Απολογισμός[γ] (€) |  |
| Προϋπολογισμός Δημοσίων Επενδύσεων | 5.812.185.821,12 |  | 6.591.651.210,23 |  |
| Μερίδιο Υπουργείου Υποδομών, Μεταφορών και Δικτύων | 1.609.700.594,37 | (27,70%) | 1.866.436.013,71 | (28,32%) |
|  | **2015** Απολογισμός (€) |  | **2016** Διαμόρφωση[γ] (€) |  |
| Προϋπολογισμός Δημοσίων Επενδύσεων | 6.377.428.124,16 |  | 6.751.114.898,00 |  |
| Μερίδιο Υπουργείου Υποδομών, Μεταφορών και Δικτύων | 1.439.541.908,12 | (22,57%) | 2.090.920.898,00 | (30,98%) |

[α] τα έτη 2011 και 2013 ο προϋπολογισμός Δημοσίων Επενδύσεων του Υπουργείου Υποδομών, Μεταφορών και Δικτύων υπήρξε κοινός με του Υπουργείου Ανάπτυξης και Ανταγωνιστικότητας και παραλείπεται
[β] Προϋπολογισμός που εκτελέστηκε
[γ] Προϋπολογισμός σε εξέλιξη
(πηγές: Βουλή των Ελλήνων[*], Υπουργείο Οικονομικών[**], προσπέλαση 23/01/2017,
[*] http://www.hellenicparliament.gr/UserFiles/2f026f42-950c-4efc-b950-340c4fb76a24/ΔΗΜΟΣΙΕΣ ΕΠΕΝΔΥΣΕΙΣ.pdf
[*] http://www.hellenicparliament.gr/UserFiles/c8827c35-4399-4fbb-8ea6-aebdc768f4f7/9391375.pdf
[**] http://www.minfin.gr/budget/2011/proyp/PDFProyp/5.1.0.pdf
[**] http://www.minfin.gr/sites/default/files/financial_files/PDE_2017.pdf)

Όπως προκύπτει από τον Πίνακα 1, οι δαπάνες Δημοσίων Επενδύσεων στην Ελλάδα που διατίθενται για τις υποδομές μεταφορών και δικτύων είναι ιδιαίτερα σημαντικές, αγγίζοντας το 1/3 του συνολικού προϋπολογισμού των Δημοσίων Επενδύσεων. Το γεγονός αυτό, σε συνδυασμό με τη δεδομένη οικονομική στενότητα που διανύει η Ελλάδα την τελευταία πενταετία (Polyzos et al., 2013b), αποδεικνύει τη βαρύτητα που δίδεται, σε επίπεδο αναπτυξιακής πολιτικής, στον τομέα των μεταφορών, ο οποίος φαίνεται πως αποτελεί βασική προοπτική για την οικονομική ανάκαμψη της χώρας, την έξοδο από την τρέχουσα οικονομική κρίση και την περαιτέρω οικονομική ανάπτυξη.

Περαιτέρω, από τον Πίνακα 2 προκύπτει ότι οι μεταφορές αντιπροσώπευαν, στη διάρκεια της περιόδου της οικονομικής ευημερίας της χώρας (1995-2004), κατά μέσο όρο το 7% του Ακαθάριστου Εθνικού Προϊόντος - ΑΕΠ (Gross National Product - GNP) της Ελλάδας (Επιλογή, 2006).

**Πίνακας 2**

α. Διαχρονική μεταβολή της ποσοστιαίας συμμετοχής των παραγωγικών τομέων στο ΑΕΠ, κατά τη δεκαετία 1995-2004

| ΠΡΟΙΟΝ/ ΕΙΣΟΔΗΜΑ[*] | 1995 | 1996 | 1997 | 1998 | 1999 | 2000 | 2001 | 2002 | 2003 | 2004 |
|---|---|---|---|---|---|---|---|---|---|---|
| ΓΕΩΡΓΙΑ[**] | 9,11 | 8,32 | 7,77 | 7,49 | 7,14 | 6,37 | 6,17 | 6,18 | 5,94 | 5,14 |

| ΠΡΟΙΟΝ/ΕΙΣΟΔΗΜΑ[(*)] | 1995 | 1996 | 1997 | 1998 | 1999 | 2000 | 2001 | 2002 | 2003 | 2004 |
|---|---|---|---|---|---|---|---|---|---|---|
| Μεταλλεία - Ορυχεία | 0,59 | 0,60 | 0,57 | 0,56 | 0,47 | 0,55 | 0,54 | 0,59 | 0,53 | 0,52 |
| Μεταποίηση | 11,97 | 12,00 | 10,68 | 10,80 | 10,61 | 10,06 | 10,02 | 9,70 | 9,58 | 9,99 |
| Ενέργεια | 2,19 | 1,97 | 1,83 | 1,89 | 1,82 | 1,55 | 1,55 | 1,55 | 1,56 | 1,61 |
| Κατασκευές | 5,94 | 5,94 | 6,04 | 6,50 | 6,57 | 6,47 | 7,34 | 7,15 | 7,62 | 7,64 |
| ΒΙΟΜΗΧΑΝΙΑ[(**)] | 20,70 | 20,51 | 19,12 | 19,75 | 19,47 | 18,64 | 19,45 | 19,00 | 19,29 | 19,77 |
| ΥΠΗΡΕΣΙΕΣ[(**)] | 62,60 | 63,00 | 64,34 | 64,24 | 63,75 | 62,25 | 61,77 | 61,99 | 62,48 | 67,04 |
| Εμπόριο | 12,53 | 13,27 | 13,17 | 12,43 | 11,96 | 11,07 | 11,87 | 11,53 | 11,62 | 12,16 |
| Ξενοδοχεία και εστιατόρια | 6,03 | 6,41 | 7,10 | 7,03 | 6,38 | 6,46 | 6,50 | 6,51 | 6,85 | 7,26 |
| *Μεταφορές-επικοινωνίες* | *6,23* | *5,68* | *5,86* | *6,14* | *7,71* | *7,61* | *7,28* | *7,46* | *7,43* | *8,65* |
| Χρηματοπιστωτική διαμεσολάβηση | 3,89 | 4,20 | 4,14 | 4,55 | 4,87 | 5,57 | 4,44 | 3,80 | 4,21 | 5,17 |

[(*)] Τιμές σε % του ΑΕΠ
[(**)] Οι βασικοί παραγωγικοί τομείς αναγράφονται με κεφαλαία
(πηγή: Επιλογή, 2006)

β. Διμεταβλητές συσχετίσεις Pearson μεταξύ των τομέων του υποπίνακα 2α

| | | Συντ. Συσχέτισης Pearson | Σημαντικότητα (δίπλευρη) | n |
|---|---|---|---|---|
| Μεταφορές - Επικοινωνίες | Α-γενής | -,855[**] | ,002 | |
| | Μεταλλεία - Ορυχεία | -,663[*] | ,037 | |
| | Μεταποίηση | -,732[*] | ,016 | |
| | Ενεργεια | -,706[*] | ,023 | |
| | Κατασκευές | ,790[**] | ,007 | |
| | Β-γενής | -,441 | ,202 | 10 |
| | Γ-γενής | ,247 | ,492 | |
| | Εμπόριο | -,740[*] | ,014 | |
| | Ξενοδοχεία - Εστιατόρια | ,187 | ,604 | |
| | Χρημ. Διαμεσολάβηση | ,564 | ,090 | |

**. Συσχέτιση σημαντική σε επίπεδο 0.01 (δίπλευρη).
*. Συσχέτιση σημαντική σε επίπεδο 0.05 (δίπλευρη).

Τα αντίστοιχα μέσα ποσοστά συμμετοχής στους βασικούς παραγωγικούς τομείς υπήρξαν 6,96% για τη γεωργία, 19,57% για την βιομηχανία και 63,34% για τις υπηρεσίες (Σχήμα 1), ενώ για τους υπόλοιπους τομείς υπήρξαν 0,55% για την εξόρυξη μεταλλευμάτων, 10,54% για την μεταποίηση, 1,75% για την ενέργεια, 6,72% για τις κατασκευές, 12,16% για το εμπόριο, 6,65% για τα ξενοδοχεία και την εστίαση και 4,48% για τη χρηματοπιστωτική διαμεσολάβηση.

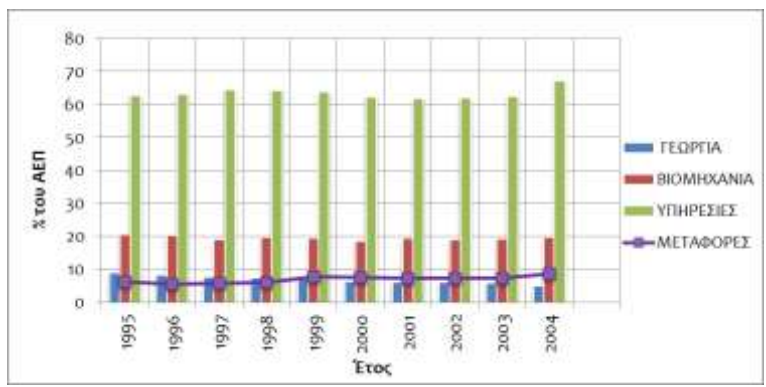

**Σχήμα 1.** Διαχρονική μεταβολή (1995-2004) της ποσοστιαίας συμμετοχής των βασικών παραγωγικών τομέων (*Α-γενής*=Γεωργία, *Β-γενής*=Βιομηχανία, *Γ-γενής*=Υπηρεσίες) και του τομέα των Μεταφορών στο ΑΕΠ (πηγή: Επιλογή, 2006; ίδια επεξεργασία).

Στο Σχήμα 1 απεικονίζεται η διαγραμματικά η διαχρονική εξέλιξη της συμμετοχής στο ΑΕΠ του τομέα των Μεταφορών - Επικοινωνιών, σε σύγκριση με την αντίστοιχη συνεισφορά των βασικών παραγωγικών τομέων της Ελλάδας (πρωτογενής – γεωργία, δευτερογενής – βιομηχανία, τριτογενής – υπηρεσίες), κατά την προαναφερόμενη περίοδο 1995-2004. Όπως προκύπτει, η δυναμική του τομέα των Μεταφορών – Επικοινωνιών καθίσταται ισοδύναμη με αυτή του πρωτογενούς τομέα, υποδηλώνοντας την εξέχουσα σημασία των μεταφορών στην οικονομική διάρθρωση της χώρας. Επιπρόσθετα, από το διάγραμμα προκύπτει ότι η δυναμική του τομέα των Μεταφορών – Επικοινωνιών στην παραγωγή του εθνικού προϊόντος είναι περίπου η μισή της αντίστοιχης του δευτερογενούς και το 10% αυτής του τριτογενούς τομέα.

Σύμφωνα με στοιχεία από την Τράπεζα της Ελλάδος (ΤτΕ, 2014), τα έσοδα από τις μεταφορές για την περίοδο 2000-2013 (Πίνακας 3), αποτέλεσαν σημαντική συνιστώσα των *εισπράξεων του τριτογενούς τομέα*, έχοντας μέσο ύψος 48%, ενώ το μέγεθός τους κάλυπτε κατά μέσο όρο το 94% του *ισοζυγίου υπηρεσιών* της χώρας. Περαιτέρω, από τα στοιχεία του Πίνακα 3 υπολογίζεται ότι τα έσοδα από μεταφορές υπερβαίνουν σε μέγεθος τα αντίστοιχα του *ταξιδιωτικού συναλλάγματος*, κατά μέσο ποσοστό 122%, αλλά και ότι υπερβαίνουν κατά μέσο ποσοστό 435% τα έσοδα από τις υπόλοιπες υπηρεσίες που καταγράφονται στο ισοζύγιο πληρωμών της χώρας.

**Πίνακας 3**
Διαχρονικά στοιχεία (2000-2013) του ισοζυγίου υπηρεσιών της χώρας και των συνιστωσών των εισπράξεων του τριτογενούς τομέα

|  | Ισοζύγιο Υπηρεσιών | Εισπράξεις Τριτογενούς Τομέα | Ταξιδιωτικό Συνάλλαγμα | Έσοδα από Μεταφορές | Λοιπές Υπηρεσίες |
|---|---|---|---|---|---|
| 2000 | 8.711,1 | 20.977 | 10.061,2 | 8.640,6 | 2.275,2 |
| 2001 | 9.150 | 22.075,9 | 10.579,9 | 9.113,3 | 2.382,7 |
| 2002 | 10.755,4 | 21.131,4 | 10.284,7 | 8.523,4 | 2.323,3 |
| 2003 | 11.506,5 | 21.430,3 | 9.495,3 | 9.569,8 | 2.365,3 |
| 2004 | 15.467 | 26.742,5 | 10.347,8 | 13.307 | 3.087,7 |
| 2005 | 15.391,1 | 27.253,5 | 10.729,5 | 13.871,4 | 2.652,6 |
| 2006 | 15.337,1 | 28.364,1 | 11.356,7 | 14.324,7 | 2.682,7 |
| 2007 | 16.591,7 | 31.337,3 | 11.319,2 | 16.939,3 | 3.078,9 |
| 2008 | 17.135,6 | 34.066,2 | 11.635,9 | 19.188,3 | 3.242 |
| 2009 | 12.640,2 | 26.983,3 | 10.400,3 | 13.552,2 | 3.030,9 |
| 2010 | 13.248,5 | 28.477,8 | 9.611,3 | 15.418,4 | 3.448,1 |
| 2011 | 14.629,6 | 28.609,2 | 10.504,7 | 14.096,6 | 4.007,9 |
| 2012 | 15.138,9 | 27.526,4 | 10.442,5 | 13.287,4 | 3.796,5 |
| 2013 | 16.978,9 | 27.959,5 | 12.152,2 | 12.089,9 | 3.717,4 |

(Πηγή: Τράπεζα της Ελλάδος, 2014; ίδια επεξεργασία)

Στο Σχήμα 2 απεικονίζονται διαγραμματικά τα στοιχεία του Πίνακα 3, συγκρίνοντας το *ταξιδιωτικό συνάλλαγμα*, τις *εισπράξεις από μεταφορές*, τις *εισπράξεις από υπηρεσίες* και το *ισοζύγιο υπηρεσιών*. Όπως προκύπτει, η δυναμική των εισπράξεων από μεταφορές παρουσιάζεται ισχυρότερη από την αντίστοιχη του ταξιδιωτικού συναλλάγματος και συγκρίσιμη με τα μεγέθη που καταγράφονται στο ισοζύγιο υπηρεσιών της χώρας.

Προκειμένου να σχηματισθεί πληρέστερη εικόνα για τη σύγκριση μεταξύ των μεγεθών *TR* (εισπράξεις από μεταφορές) – *TE* (ταξιδιωτικό συνάλλαγμα), *TR* (εισπράξεις από μεταφορές) – *SB* (ισοζύγιο υπηρεσιών) και *TE* – *SB*, εφαρμόζεται η στατιστική δοκιμασία κατά ζεύγη *paired-samples t*-test (Norusis, 2004) για τον έλεγχο της μηδενικής υπόθεσης $h_o$: $μ_i-μ_j=0$ (μηδενική υπόθεση), έναντι της εναλλακτικής $h_1$: $μ_i-μ_j≠0$. Η διαδικασία υπολογίζει τις διαφορές μεταξύ των τιμών των δύο μεταβλητών για κάθε περίπτωση και εξετάζει εάν ο μέσος όρος τους διαφέρει στατιστικά από μηδέν. Σε αυτήν τη δοκιμασία επιλέγεται διάστημα εμπιστοσύνης 95% για τη διαφορά των μέσων. Τα διαθέσιμα στοιχεία διαχειρίζονται ζευγαρωτά (*pair-wise*), υπονοώντας ότι κάθε δοκιμασία *t*-test υπολογίζεται ανά περίπτωση στο μεγαλύτερο δυνατό αριθμό δείγματος (βαθμών ελευθερίας) που μπορεί να ποικίλει από δοκιμασία σε δοκιμασία.

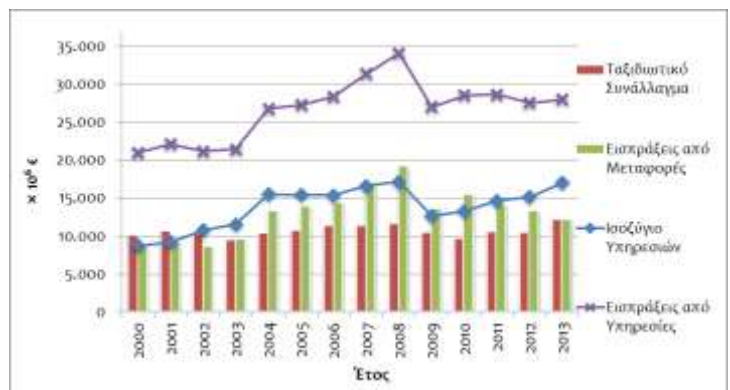

**Σχήμα 2.** Διαχρονική μεταβολή (2000-2013) του Ταξιδιωτικού Συναλλάγματος και των εισπράξεων από Μεταφορές, σε σύγκριση με τις Εισπράξεις από Υπηρεσίες και του Ισοζυγίου Υπηρεσιών (πηγή: Τράπεζα της Ελλάδος, 2014; ίδια επεξεργασία).

Τα αποτελέσματα της στατιστικής δοκιμασίας (στατιστικού ελέγχου) κατά ζεύγη παρουσιάζονται στον Πίνακα 4, όπου προκύπτει πως οι διαφορές *TR-TE* και *TE-SB* είναι στατιστικά σημαντικές, ενώ η *TR-SB* όχι. Η στατιστικά σημαντική διαφορά *TR-TE*>0 εκφράζει ότι κατά την περίοδο 2000-2013 η μέση τιμή των εισπράξεων από μεταφορές είναι στατιστικά μεγαλύτερη της αντίστοιχης του τουριστικού συναλλάγματος, ενώ η στατιστικά μηδενική διαφορά *TR-SB*=0 ότι η μέση τιμή των εισπράξεων από μεταφορές δεν διαφέρει από αυτή του ισοζυγίου υπηρεσιών. Αντίθετα, η στατιστικά σημαντική διαφορά *TE-SB*<0 εκφράζει ότι, την εν λόγω περίοδο, η μέση τιμή του τουριστικού συναλλάγματος υπήρξε μικρότερη από την αντίστοιχη του ισοζυγίου υπηρεσιών.

Τα αποτελέσματα της στατιστικής δοκιμασίας (στατιστικού ελέγχου) κατά ζεύγη οδηγούν στο γενικό συμπέρασμα ότι ο τομέας των μεταφορών είχε μεγαλύτερο αντίκτυπο στην εθνική οικονομία σε σχέση με τα έσοδα από τουριστικό συνάλλαγμα, κατά την περίοδο 2000-2013, αλλά και πως το μέγεθός του είναι αντίστοιχο με αυτό του ισοζυγίου υπηρεσιών της χώρας.

**Πίνακας 4**
Στατιστική δοκιμασία κατά ζεύγη (paired-samples *t*-test) για τη σύγκριση των μέσων τιμών $μ_i$ και $μ_j$, ανάμεσα στις μεταβλητές του ισοζυγίου υπηρεσιών.

4α. Σύνοψη

| | Ζεύγος | β.ε | Συσχετίσεις ζευγών $r_{xy}$ | Σημ[α] |
|---|---|---|---|---|
| 1 | TR[β]-TE[γ] | 14 | 0,449 | 0,108 |
| 2 | TR-SB[δ] | 14 | 0,809 | 0,000 |

| Ζεύγος | | Διαφορές ζευγών | | | | | | | |
|---|---|---|---|---|---|---|---|---|---|
| | | Μέσος | s | $s_e^{(στ)}$ του μέσου | 95% δ.ε$^{(ε)}$ για τη διαφορά μέσων | | t | β.ε | Σημ.$^{(b)}$ |
| | | | | | Κατώτ. | Ανώτερο | | | |
| 3 | TE-SB | 14 | 0,664 | 0,010 | | | | | |

4β. Αποτελέσματα δοκιμασίας

| | Ζεύγος | Μέσος | s | $s_e^{(στ)}$ του μέσου | Κατώτ. | Ανώτερο | t | β.ε | Σημ.$^{(b)}$ |
|---|---|---|---|---|---|---|---|---|---|
| 1 | TR$^{(d)}$–TE$^{(e)}$ | 2357,2 | 2911,56 | 778,15 | **676,14** | **4038,3** | 3,029 | 13 | ,010 |
| 2 | TR–SB$^{(f)}$ | -768,5 | 1880,25 | 502,52 | -1854,1 | 317,10 | -1,529 | 13 | ,150 |
| 3 | TE–SB | -3125,7 | 2382,48 | 636,74 | **-4501,3** | **-1750,1** | -4,909 | 13 | ,000 |

α. Δίπλευρη σημαντικότητα  
β. Έσοδα από μεταφορές  
γ. Τουριστικό συνάλλαγμα  
δ. Έσοδα από υπηρεσίες  
στ. Τυπικό σφάλμα  
ε. (δ.ε) = Διάστημα εμπιστοσύνης  
(πηγή δεδομένων: Τράπεζα της Ελλάδος, 2014)

Κατασκευάζοντας τα διαγράμματα διασποράς μεταξύ των μεταβλητών *TR* (εισπράξεις από μεταφορές) – *SI* (έσοδα τομέα υπηρεσιών) (Σχήμα 3) και *TR – SB* (ισοζύγιο υπηρεσιών) (Σχήμα 4), διαφαίνεται πως υφίσταται ισχυρότερη δομική σχέση μεταξύ των μεταβλητών *TR – SI* σε σχέση με τις μεταβλητές *TR – SB*.

Ειδικότερα, η παρατήρηση αυτή εκφράζει ότι το σύνολο σχεδόν (95,3%) της διαχρονικής διακύμανσης των στοιχείων των εσόδων από μεταφορές περιγράφεται επίσης από τη διακύμανση των στοιχείων των εσόδων του τριτογενούς τομέα, ενώ στην περίπτωση της διαφοράς *TR – SB* η ικανότητα αυτή πέφτει στο 65,4%.

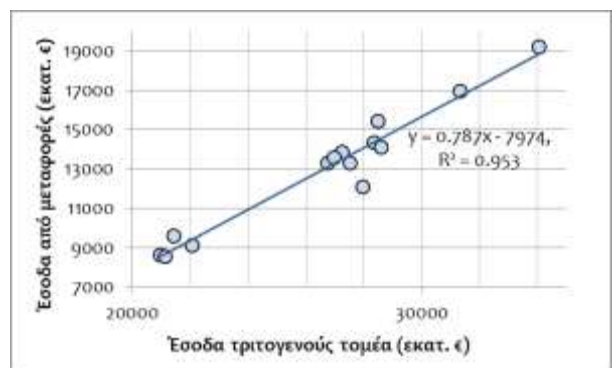

**Σχήμα 3.** Διαχρονική μεταβολή (2000-2013) των εισπράξεων από Μεταφορές σε σύγκριση με τις Εισπράξεις από Υπηρεσίες (πηγή: Τράπεζα της Ελλάδος, 2014; ίδια επεξεργασία).

Για την περαιτέρω ανίχνευση της δομικής συνάφειας στη μεταβλητότητα των μεγεθών του Πίνακα 3 πραγματοποιείται *ανάλυση συσχετίσεων* (*correlation analysis*) (Μαχαίρα και Μπόρα, 1998; Norusis, 2004; Devore and Berk, 2012), με χρήση του διμετάβλητου συντελεστή γραμμικής συσχέτισης $r_{xy}$ του Pearson. Ο εν λόγω συντελεστής ανιχνεύει την ύπαρξη γραμμικής σχέσης μεταξύ δύο μεταβλητών *x,y* και οι τιμές του κυμαίνονται στο διάστημα [-1,1]. Στην περίπτωση $r_{xy}$=1 υφίσταται πλήρης θετική γραμμική σχέση μεταξύ των μεταβλητών *x,y*, όταν $r_{xy}$=-1 πλήρης αρνητική γραμμική σχέση, ενώ όταν $r_{xy}$=0 οι μεταβλητές είναι γραμμικά ασυσχέτιστες. Τα αποτελέσματα της ανάλυσης συσχέτισης παρουσιάζονται στον Πίνακα 5.

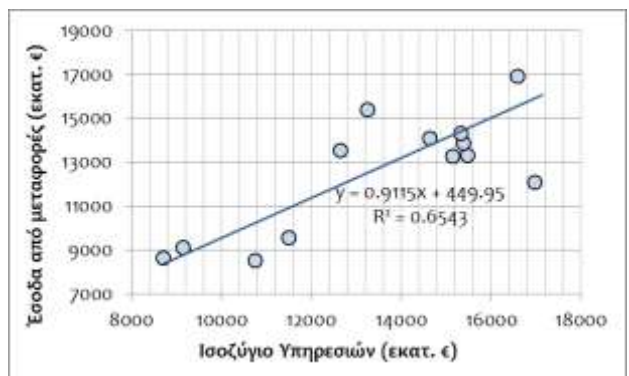

**Σχήμα 4.** Διαχρονική μεταβολή (2000-2013) των εισπράξεων από Μεταφορές σε σύγκριση με το Ισοζύγιο Υπηρεσιών (δεξ.) (πηγή: Τράπεζα της Ελλάδος, 2014; ίδια επεξεργασία).

Από τον Πίνακα 5 προκύπτει ότι τα έσοδα από τις μεταφορές (*TR*) είναι σχεδόν πλήρως γραμμικά συσχετισμένα με τα έσοδα του τριτογενούς τομέα (*SI*), έχοντας συντελεστή συσχέτισης $r_{TR,SI}=0{,}976^{**}$ με σημαντικότητα $p_{TR,SI}\sim 0$, αλλά και πολύ συσχετισμένα με το ισοζύγιο υπηρεσιών, όπου $r_{TR,SB}=0{,}809^{**}$ με σημαντικότητα επίσης $p_{TR,SI}\sim 0$.

**Πίνακας 5**
Ανάλυση συσχετίσεων για τα δεδομένα του Πίνακα 3

|    |              | SB[d] | SI[e]     | TE[f]    | TR[g]     | OS[h]     |
|----|--------------|-------|-----------|----------|-----------|-----------|
| SB | $r_{xy}$[a]  | 1     | 0,877**   | 0,664**  | 0,809**   | 0,661**   |
|    | Sig.[b]      |       | 0,000     | 0,010    | 0,000     | 0,010     |
|    | N[c]         | 14    | 14        | 14       | 14        | 14        |
| SI | $r_{xy}$     |       | 1         | 0,598*   | 0,976**   | 0,665**   |
|    | Sig.         |       |           | 0,024    | 0,000     | 0,009     |
|    | N            |       | 14        | 14       | 14        | 14        |
| TE | $r_{xy}$     |       |           | 1        | 0,449     | 0,309     |
|    | Sig.         |       |           |          | 0,108     | 0,282     |
|    | N            |       |           | 14       | 14        | 14        |
| TR | $r_{xy}$     |       |           |          | 1         | 0,568**   |
|    | Sig.         |       |           |          |           | 0,034     |
|    | N            |       |           |          | 14        | 14        |

a. Συντελεστής συσχέτισης του Pearson
b. Σημαντικότητα (*p*-value)
c. Βαθμοί ελευθερίας
d. Ισοζύγιο Υπηρεσιών
e. Εισπράξεις Τριτογενούς Τομέα
f. Ταξιδιωτικό Συνάλλαγμα
g. Έσοδα από Μεταφορές
h. Λοιπές Υπηρεσίες
\**. Η συσχέτιση είναι σημαντική σε επίπεδο 0.01 (2-πλευρη).
\*. Η συσχέτιση είναι σημαντική σε επίπεδο 0.05 (2-πλευρη).
(πηγή: Τράπεζα της Ελλάδος, 2014; Ίδια επεξεργασία)

Οι αντίστοιχες τιμές για το ταξιδιωτικό συνάλλαγμα (*TE*) και τα έσοδα από λοιπές υπηρεσίες (*OS*) είναι αισθητά μικρότερες από τις προαναφερόμενες, της τάξεως του 20-40%.

Τα προαναφερόμενα αποτελέσματα εκφράζουν ότι η διαχρονική μεταβλητότητα στις τιμές των εσόδων του τριτογενούς τομέα (*SI*) αποτυπώνεται με την ίδια σχεδόν αναλογία και στη διαχρονική εξέλιξη των τιμών των εσόδων από τις μεταφορές (*TR*), αναδεικνύοντας την ύπαρξη δομικής συνάφειας μεταξύ των μεταβλητών *SI* και *TR*.

Το ίδιο συμβαίνει, αλλά ελαφρώς ασθενέστερα, με τις μεταβλητές *SB* και *TR*. Η παρατήρηση αυτή, σε σχέση με τα μικρότερα αποτελέσματα των μεταβλητών *TE* και *OS* αντίστοιχα, υποδηλώνει την ισχυρότερη συμμετοχή των εσόδων από μεταφορές στη διαμόρφωση της μεταβλητότητας των εσόδων του τριτογενούς τομέα και του ισοζυγίου υπηρεσιών.

### 3. Τα δίκτυα μεταφορών στην Ελλάδα

Όπως προέκυψε από την παρουσίαση και στατιστική επεξεργασία των στοιχείων που αντλήθηκαν τόσο από τη διοικητική όσο και από την οικονομική δομή της χώρας, ο τομέας των μεταφορών στην Ελλάδα φαίνεται πως αποτελεί βασική συνιστώσα ανάπτυξης. Στο πλαίσιο αυτό, στις υποενότητες που ακολουθούν εξειδικεύεται η μελέτη των μεταφορών στην περίπτωση της Ελλάδας, αποκτώντας περισσότερο τεχνική βάση και εστιάζοντας στη δομή των δικτύων μεταφορών.

Τα δίκτυα που παρουσιάζονται παρακάτω είναι το *οδικό*, το *σιδηροδρομικό*, το *ακτοπλοϊκό* και το *αεροπορικό* δίκτυο της χώρας.

*3.1. Το οδικό δίκτυο μεταφορών*
3.1.1. Ιστορικό πλαίσιο
Η διαχρονική εξέλιξη των οδικών δικτύων μεταφορών είναι συνυφασμένη με την ιστορία της ανθρωπότητας. Αρχικά οι ανθρώπινες μετακινήσεις πραγματοποιούνταν με αξιοποίηση των φυσικών διαύλων, ακολουθώντας δηλαδή την πορεία των ποταμών ή άλλων φυσικών διαβάσεων. Ωστόσο, ιστορικό κριτήριο της εμφάνισης των οδικών μεταφορών αποτέλεσε η εκμετάλλευση της ανθρώπινης σκέψης και της ανθρώπινης εργασίας στο σχεδιασμό και την κατασκευή και γενικά στην ανάπτυξη των οδικών υποδομών (Καλτσούνης, 2007). Σύμφωνα με τον Καλτσούνη (2007), στην ιστορική αναδρομή της οδοποιίας επισημαίνονται τα εξής:

• Πρώτες ενδείξεις αξιοποίησης της ανθρώπινης σκέψης και εργασίας στο σχεδιασμό και κατασκευή οδικού δικτύου καταγράφονται οι λιθόστρωτες οδοί που κατασκευάσθηκαν στη Μεσοποταμία, οι οποίες χρονολογούνται γύρω στο 4000 π.Χ. (την περίοδο ανακάλυψης του τροχού). Ακολουθούν χρονολογικά οι πλινθόστρωτες οδοί στην Ινδία (γύρω στο 3000 π.Χ.) και οι λιθόστρωτοι δρόμοι της Μινωικής Εποχής στην Κρήτη.

• Η αρχαιότερη οδός που διατηρείται μέχρι σήμερα βρίσκεται στην Κρήτη, κατασκευάστηκε γύρω στο 1700π.Χ. και έχει συνολικό μήκος περίπου 50km. Η οδός ένωνε την Κνωσό με την πόλη Γόρτυνα και τις νότιες ακτές της Νήσου.

• Στην εποχή της Αρχαίας Ελλάδας τέθηκαν οι βάσεις του αστικού οδικού σχεδιασμού. Η αστική οδός την περίοδο αυτή έπαψε να ακολουθεί τις διαμορφωμένες τυχαίες εδαφικές χαράξεις, αποτελώντας προϊόν σχεδιασμού, προσαρμοσμένο στις οικιστικές ανάγκες, το οποίο συνθέτει ένα οργανωμένο οδικό δίκτυο. Την εποχή αυτή εμφανίζεται η ρυμοτομική δράση της πόλης σε οικοδομικά τετράγωνα, κατά την οποία λαμβάνεται μέριμνα για το σχεδιασμό των δημόσιων χώρων και των απαιτούμενων οδικών υποδομών, με χαρακτηριστικό αντιπρόσωπο την πόλη της Μιλήτου. Κυρίαρχο όχημα της περιόδου αυτής υπήρξαν οι τροχήλατες ιππήλατες άμαξες.

• Η συμβολή του Μ. Αλεξάνδρου στην οδοποιία της Αρχαίας Ελλάδας υπήρξε σημαντική, καθόσον συντήρησε, βελτίωσε και επέκτεινε, με τους Θρακιώτες τεχνίτες του, το οδικό δίκτυο των Περσών που παρέλαβε κατά τη διάρκεια της αυτοκρατορίας του.

• Καθοριστική υπήρξε η εξέλιξη της οδοποιίας κατά την ρωμαϊκή εποχή, η οποία διευθύνθηκε από την ανάγκη ελέγχου αυτής της αχανούς αυτοκρατορίας, καθόσον το οδικό δίκτυό της εκτείνονταν από τη Βόρεια Θάλασσα μέχρι τη Σαχάρα και από τον Ατλαντικό μέχρι τη Μεσοποταμία, με κομβικό σημείο τη Ρώμη. Η αρτιότητα στην κατασκευή του εν λόγω δικτύου επέτρεψε τη λειτουργία του για πολλούς αιώνες έπειτα από την κατάρρευση της ρωμαϊκής αυτοκρατορίας. Βασικό στοιχείο του οδικού δικτύου εκείνης της εποχής υπήρξε η ιεράρχηση στην κατασκευή των τμημάτων του, του οποίου το πρωτεύον τμήμα ήταν λιθόστρωτο (μήκους 90.000km), το δευτερεύον χαλικόστρωτο (μήκους 300.000km) και τα υπόλοιπα τμήματα ξυλόστρωτα (σανιδόστρωτα). Αξιοσημείωτο υπήρξε το γεγονός ότι το υπόψη δίκτυο έφερε πληροφοριακή σήμανση, υπήρξε χιλιομετρημένο, συντηρούνταν συστηματικά και διέθετε υποδομές στάσης και διανυκτέρευσης (ξενώνες). Χαρακτηριστικό γνώρισμα του δικτύου υπήρξε η εκτενής ευθυγράμμισή του (οι λεγόμενες «*Ρωμαϊκές Ευθυγραμμίες*»), η οποία εξυπηρετούσε λόγους ασφάλειας και μείωσης των αποστάσεων και παρακάμπτονταν μόνο για την καλύτερη προσαρμογή του έργου στο έδαφος. Το εν λόγω δίκτυο χαρακτηρίζονταν επίσης από συχνές κατά μήκος κλίσεις, οι οποίες ανέρχονταν στο 10% και ισχύουν και σήμερα ως επιτρεπτές κλίσεις στον οδικό σχεδιασμό, για ταχύτητα μελέτης μικρότερη των 60km/*h*.

- Σταθμό στην εξέλιξη της οδοποιίας αποτέλεσε η εφεύρεση του αυτοκινήτου, η οποία συντέλεσε στην αναβάθμιση τόσο του γεωμετρικού σχεδιασμού όσο και της ποιότητας κατασκευής των οδών. Πριν από την εμφάνιση του αυτοκινήτου δεν υπήρχαν αυστηρές απαιτήσεις στη γεωμετρική χάραξη των οδών, λόγω των μικρών ταχυτήτων μετακίνησης των ιππήλατων και λοιπών ζωήλατων μέσων. Η εμφάνιση όμως του αυτοκινήτου σηματοδότησε μια νέα εποχή στις μετακινήσεις που πραγματοποιούνταν με συνεχώς αυξανόμενες ταχύτητες. Στόχο της οδοποιίας αποτέλεσε έκτοτε η κατασκευή οδικών δικτύων ποιοτικών υποδομών, με πρωτεύον μέλημα την ασφάλεια των χρηστών.

Στην Ελλάδα η κατασκευή του οδικού δικτύου στη σημερινή του μορφή ξεκίνησε επίσης μεταπολεμικά. Το 1963, με Απόφαση του Υπουργού Δημοσίων Έργων (Γ.25871, 1963) καταρτίστηκε κατάλογος με την αρίθμηση των εθνικών οδών της χώρας (βλ. Παράρτημα), με σκοπό «*τη συστηματική χιλιομέτρησή τους και την καλύτερη εξυπηρέτηση της αυξανόμενης τουριστικής κίνησης της χώρας*», στον οποίο περιλαμβάνονταν και τμήματα που δεν είχαν ακόμη ασφαλτοστρωθεί. Μέσα από τη διατύπωση του σκοπού της παραπάνω απόφασης εκφράζεται - ήδη από το 1963 - η νομοθετική και πολιτική αντίληψη, αναφορικά με τη σχέση αλληλεπίδρασης μεταξύ των μεταφορών και του τουρισμού και την αναμενόμενη συμβολή τους στην οικονομική ανάπτυξη της χώρας. Το 2006 η Γενική Γραμματεία Δημοσίων Έργων (ΓΓΔΕ) του Υπουργείου Περιβάλλοντος Χωροταξίας και Δημοσίων Έργων προέβη στη σύνταξη του «*Οδηγού Χιλιομετρικών Αποστάσεων του Οδικού Δικτύου της Χώρας*», με σκοπό την εξυπηρέτηση των σχετικών αναγκών των υπαλλήλων και των ιδιωτών, τη διευκόλυνση του έργου των Υπηρεσιών Δημοσιονομικού Ελέγχου των Υπουργείων και γενικά των Δημόσιων Υπηρεσιών, αλλά και τη διευκόλυνση έκδοσης βεβαιώσεων χιλιομετρικών αποστάσεων μεταξύ Δήμων και Κοινοτήτων της χώρας. Ο οδηγός περιλαμβάνει πίνακες με τις χιλιομετρικές αποστάσεις των πρωτευουσών των Καποδιστριακών νομών της χώρας μεταξύ τους, από την Αθήνα και τις αποστάσεων πόλεων, χωριών και οικισμών κάθε νομού από την πρωτεύουσά του (ΥΠΕΧΩΔΕ, 2006).

3.1.2. Τεχνικά στοιχεία
Τα οδικά δίκτυα μεταφορών ανήκουν στην οικογένεια των *δικτύων υποδομών* (*infrastructure networks*) (Barthelemy, 2011), διότι οι οδικές συνδέσεις που τα απαρτίζουν συντελούνται πάνω σε δομημένες επιφάνειες και συνιστούν έργα εθνικής υποδομής. Σύμφωνα με το ΠΔ.401/93 το οδικό δίκτυο κατατάσσεται στις παρακάτω κατηγορίες, με κύρια κριτήρια τα εξής:

- *Βασικό εθνικό οδικό δίκτυο*: αποτελεί το τμήμα εκείνο του εθνικού οδικού δικτύου που συνδέει τα σπουδαιότερα αστικά κέντρα μεταξύ τους και τη χώρα με άλλες επικράτειες, είτε απευθείας ή με παρέμβαση πορθμείων.
- *Δευτερεύον εθνικό οδικό δίκτυο*: ονομάζεται το τμήμα του εθνικού οδικού δικτύου που συνδέει βασικούς εθνικούς οδικούς άξονες μεταξύ τους ή με μεγάλα αστικά κέντρα, λιμάνια, αεροδρόμια ή με τόπους εξαιρετικού τουριστικού ενδιαφέροντος ή είναι οδικοί άξονες για τους οποίους έχει γίνει παραλλαγή με Βασικό Εθνικό Οδικό Δίκτυο.

- *Τριτεύον εθνικό οδικό δίκτυο*: ονομάζεται το τμήμα του εθνικού οδικού δικτύου που έχει αντικατασταθεί με νέες χαράξεις Εθνικού Οδικού Δικτύου ή εξυπηρετεί μετακινήσεις σε περιοχές με αρχαιολογικό τουριστικό, ιστορικό ή αναπτυξιακό ενδιαφέρον.
- *Πρωτεύον επαρχιακό οδικό δίκτυο*: αποτελεί το τμήμα του επαρχιακού οδικού δικτύου που συνδέει αστικά κέντρα με το εθνικό οδικό δίκτυο, καθώς και περιοχές με αρχαιολογικό, τουριστικό, ιστορικό ή αναπτυξιακό ενδιαφέρον.
- *Δευτερεύον επαρχιακό οδικό δίκτυο*: είναι το τμήμα εκείνο του επαρχιακού οδικού δικτύου που συνδέει Δήμους ή Κοινότητες εκτός της Πρωτεύουσας του Νομού μεταξύ τους.

Στο ίδιο προεδρικό διάταγμα (ΠΔ.401/93), η κατάταξη του εθνικού οδικού δικτύου σε βασικό, δευτερεύον και τριτεύον δίκτυο γίνεται σύμφωνα με τα παραπάνω κριτήρια και με Απόφαση του Υπουργού Περιβάλλοντος, Χωροταξίας και Δημόσιων Έργων, μετά από σύμφωνη γνώμη του Συμβουλίου Δημοσίων Έργων και δημοσιεύεται στην Εφημερίδα της Κυβερνήσεως. Επίσης, η κατάταξη του επαρχιακού οδικού δικτύου σε πρωτεύον και δευτερεύον γίνεται με Απόφαση του Υπουργού Περιβάλλοντος, Χωροταξίας και Δημοσίων Έργων και εισήγηση του Γενικού Γραμματέα Περιφέρειας μετά από πρόταση των κατά τόπους αρμοδίων Νομαρχών και σύμφωνη γνώμη του Νομαρχιακού Συμβουλίου Δημόσιων Έργων και δημοσιεύεται στην Εφημερίδα της Κυβέρνησης.

Σύμφωνα με τα διατιθέμενα στοιχεία στη βάση δεδομένων του Οργανισμού Κτηματολογίου και Χαρτογραφήσεως Ελλάδας (ΟΚΧΕ, 2015), το εθνικό και επαρχιακό οδικό δίκτυο της Ελλάδας έχει συνολικό μήκος 35.860km και εξυπηρετεί το σύνολο των 51 πρωτευουσών των ελληνικών καποδιστριακών νομών. Στο Σχήμα 5 απεικονίζεται η χαρτογράφηση του υπόψη δικτύου, μαζί με τις 51 πρωτεύουσες των νομών. Σήμερα η χώρα μας διαθέτει περίπου 1900 χιλιόμετρα αυτοκινητοδρόμων ενώ με την ολοκλήρωση των τμημάτων που βρίσκονται υπό κατασκευή αναμένεται να ξεπεράσει τα 3.100km (ΥΠΕΧΩΔΕ, 2006).

Οι βασικοί οδικοί άξονες στην Ελλάδα (βλ. Παράρτημα), όπως προκύπτει από το Στρατηγικό Σχέδιο Ανάπτυξης της Συγκοινωνιακής Υποδομής στην Ελλάδα για το έτος 2010 (ΥΠΕΘΟ, 1993; Εγνατία Οδός ΑΕ, 2008) και την κωδικοποίηση που πραγματοποιήθηκε με την Υπουργική Απόφαση, με αριθμό πρωτοκόλλου ΔΜΕΟ/ο/7157/ε/1042, του Υπουργού Περιβάλλοντος Χωροταξίας και Δημοσίων Έργων (ΥΠΕΧΩΔΕ, 2008), παρουσιάζονται στον Πίνακα 6.

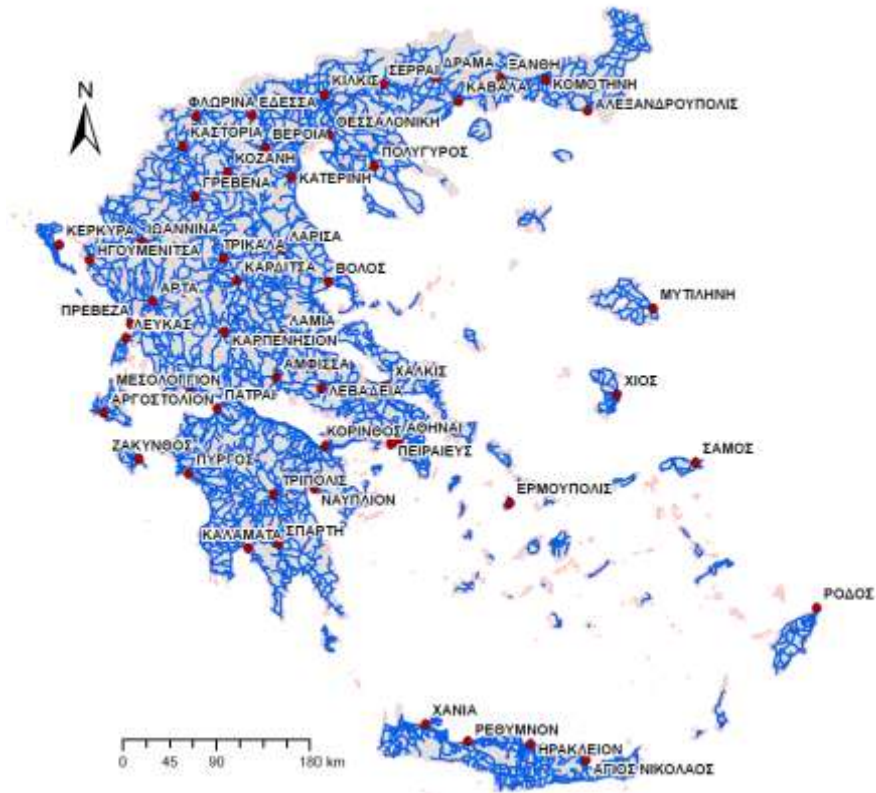

**Σχήμα 5.** Το εθνικό και επαρχιακό δίκτυο οδικών μεταφορών της Ελλάδας (πηγή βάσης δεδομένων: ΟΚΧΕ, 2005).

Ειδικότερα, οι βασικοί οδικοί άξονες της χώρας (βλ. Παράρτημα) είναι οι εξής (πηγές: ΥΠΕΘΟ, 1993; Εγνατία Οδός ΑΕ, 2008; ΥΠΕΧΩΔΕ, 2006, 2008, Polyzos et al., 2014):

• Ο *αυτοκινητόδρομος Πάτρα – Αθήνα – Θεσσαλονίκη - Εύζωνοι (ΠΑΘΕ)*: Έχει συνολικό μήκος 770km και αποτελεί το βασικό οδικό άξονα της ηπειρωτικής Ελλάδας, διατρέχοντας έξι περιφέρειες, την *Κεντρική Μακεδονία*, τη *Θεσσαλία*, τη *Στερεά Ελλάδα*, την *Αττική*, την *Πελοπόννησο* και τη *Δυτική Ελλάδα* (βλ. Παράρτημα). Ο κωδικός που έχει λάβει, σύμφωνα με την «Κωδικοποίηση και αρίθμηση του Ελληνικού Διευρωπαϊκού Οδικού Δικτύου» (ΥΠΕΘΑ, 2008), είναι ο Α1 (Αθήνα-Θεσσαλονίκη-Εύζωνοι) και ο Α8 (Πάτρα-Αθήνα). Ο ΠΑΘΕ συνδέει τις τρεις πολυπληθέστερες πόλεις της Ελλάδας (Αθήνα, Θεσσαλονίκη, Πάτρα) και κατά μήκος του διαρθρώνεται ένα πλέγμα από τα σημαντικότερα αστικά κέντρα και οικισμούς της χώρας, με πληθυσμό που ξεπερνά το 50% του συνολικού πληθυσμού. Ο αυτοκινητόδρομος έχει κατασκευαστεί στο μεγαλύτερο μέρος του, με εξαίρεση δύο μικρά τμήματα στις περιοχές του Μαλιακού και των Τεμπών, τα οποία βρίσκονται σε εξέλιξη.

• Ο *αυτοκινητόδρομος της Εγνατίας Οδού*: Διατάσσεται σχεδόν κάθετα προς τον ΠΑΘΕ, έχει κωδικό Α2 (ΥΠΕΘΑ, 2008), συνολικό μήκος 670km και κοινή διαδρομή με την Α1 μήκους 26km. Διατρέχει πέντε περιφέρειες της βόρειας Ελλάδας, την *Ανατολική Μακεδονία και Θράκη*, την *Κεντρική Μακεδονία*, τη *Δυτική Μακεδονία*, τη *Θεσσαλία* (σε ένα μικρό τμήμα 15km στο βόρειο τμήμα του νομού Τρικάλων) και την *Ήπειρο* (βλ. Παράρτημα). Η κατασκευή του αυτοκινητόδρομου της Εγνατίας Οδού ξεκίνησε το έτος 1997, με εξαίρεση των κοινών 26km με τον ΠΑΘΕ, τα οποία κατασκευάστηκαν πριν το 1994, και σήμερα το έργο είναι ολοκληρωμένο. Η Εγνατία Οδός έχει αφετηρία την Ηγουμενίτσα και απόληξη τις εξόδους Κήπων και Καστανιών (προς Τουρκία) και διατρέχει όλο το γεωγραφικό διαμέρισμα της Βόρειας Ελλάδας, καθώς και την περιφέρεια της Ηπείρου. Ως έργο έχει ιδιαίτερη σημασία για την Βόρεια Ελλάδα, συνδέοντας τα περισσότερα αστικά κέντρα του βορειοελλαδικού χώρου.

- Ο *αυτοκινητόδρομος Δυτικής Ελλάδας* (*Ιόνια οδός*): Έχει κωδικό Α5 (ΥΠΕΘΑ, 2008), αφετηρία την Καλαμάτα και η χάραξή της διατρέχει κατά μήκος το γεωγραφικό διαμέρισμα της Δυτικής Ελλάδας, διά μέσου της ζεύξης Ρίου – Αντιρρίου (που αποτελεί ένα από τα σημαντικότερα έργα υποδομής στην Ευρώπη), καταλήγοντας στα Ελληνοαλβανικά σύνορα. Η Ιόνια Οδός διατρέχει στο σύνολό της τρεις Περιφέρειες, την *Πελοπόννησο*, τη *Δυτική Ελλάδα* και την *Ήπειρο* (βλ. Παράρτημα), με μήκος που πρόκειται να ανέρχεται, έπειτα από την ολοκλήρωσή της, σε 375km. Το έργο βρίσκεται στο μεγαλύτερο μέρος του σε εξέλιξη, με περατωμένο μέρη μόνο τη ζεύξη Ρίου – Αντιρρίου και δύο τμημάτων που διατρέχουν τους νομούς Αιτωλοακαρνανίας και Πρεβέζης. Η κατασκευή της Ιόνιας Οδού αναμένεται να αποτελέσει τη ραχοκοκαλιά της συγκοινωνιακής υποδομής της Δυτικής Ελλάδας, αναβαθμίζοντας την ποιότητα των συγκοινωνιών και προάγοντας την περιφερειακή ανάπτυξη.
- Ο *αυτοκινητόδρομος Κεντρικής Ελλάδας* (*Ε65*): Έχει κωδικό Α3 (ΥΠΕΘΑ, 2008) και χάραξη συνολικού μήκους 175km. Η όδευσή του εκτείνεται από τη Λαμία ως την Εγνατία Οδό, με την οποία συνδέεται στο ύψος του Μετσόβου (βλ. Παράρτημα). Το έργο βρίσκεται υπό μελέτη και αποτελείται από δύο τμήματα, το πρώτο από την Παναγιά Τρικάλων μέχρι τη Λαμία (διασχίζοντας Καλαμπάκα, Τρίκαλα και Καρδίτσα) και το δεύτερο από τη Λαμία μέχρι το Αντίρριο (μέσω Άμφισσας και Ιτέας). Ο αυτοκινητόδρομος Κεντρικής Ελλάδας πρόκειται να διέρχεται τις περιφέρειες της *Δυτικής Μακεδονίας*, *Θεσσαλίας* και *Στερεάς Ελλάδας* και θα επεκτείνεται στην *Περιφέρεια Δυτικής Ελλάδας*.
- Ο *Κεντρικός Άξονας Πελοποννήσου*: Εξυπηρετεί αποκλειστικά την περιφέρεια της Πελοποννήσου έχοντας συνολικό μήκος περίπου 200km και η κατασκευή του βρίσκεται σε εξέλιξη. Έχει κωδικό Α7 (ΥΠΕΘΑ, 2008), αφετηρία την Κόρινθο και εκτείνεται μέχρι την Καλαμάτα, διακλαδώνοντας κάθετα μέχρι τη Σπάρτη (βλ. Παράρτημα).
- Ο *Βόρειος Άξονας της Κρήτης* έχει κωδικό Α90 (ΥΠΕΘΑ, 2008), συνολικό μήκος περίπου 300km και η κατασκευή του βρίσκεται σε εξέλιξη, έχοντας περατωθεί το μεγαλύτερο μέρος του (περισσότερα από 200km). Ο άξονας καλύπτει το βόρειο τμήμα της Κρήτης, δυτικά από το Καστέλι μέχρι ανατολικά στη Σητεία, διατρέχοντας και τις τέσσερεις πρωτεύουσες των Καποδιστριακών νομών του νησιού, τα *Χανιά*, το *Ηράκλειο*, τον *Άγιο Νικόλαο* και το *Ρέθυμνο*.

Ως προς το λειτουργικό του φόρτο, το οδικό δίκτυο της Ελλάδας είναι επιφορτισμένο με ένα μεγάλο αριθμό οχημάτων, ο οποίος βαίνει αυξανόμενος διαχρονικά. Το διάγραμμα του σχήματος 6 παρουσιάζει τη χρονική εξέλιξη του αριθμού των καταγεγραμμένων προς κυκλοφορία οχημάτων στη χώρα, κατά την περίοδο 1990-2010, καθώς και τον εκτιμώμενο αριθμό επιβατικών οχημάτων μέχρι το 2020 (ΥΠΕΚΑ, 2011; ΕΛΣΤΑΤ, 2018). Λαμβάνοντας υπόψη το διάγραμμα του σχήματος 6 και το συνολικό μήκος του οδικού δικτύου της χώρας, όπως προκύπτει από το χάρτη του σχήματος 5, υπολογίζεται ο λόγος *αριθμός οχημάτων/μήκος οδικού δικτύου* που αποτελεί μέγεθος γραμμικής πυκνότητας του ελληνικού οδικού δικτύου.

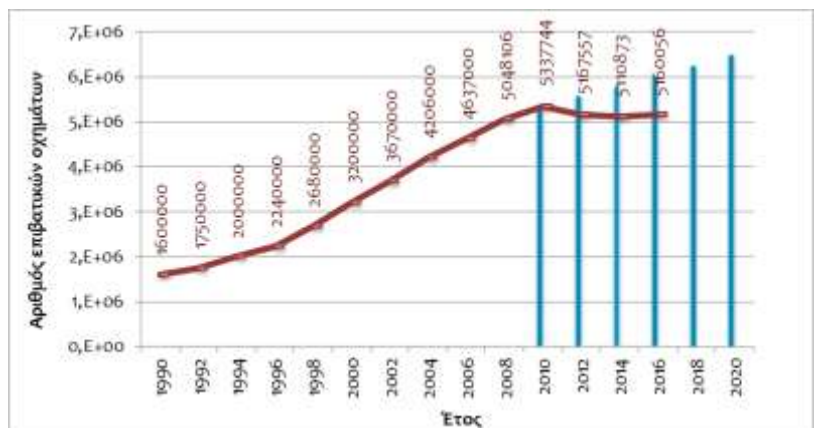

**Σχήμα 6.** Αριθμός επιβατικών οχημάτων που κυκλοφορούν στην Ελλάδα. Οι τιμές των ράβδων αποτελούν πρόβλεψη του ΥΠΕΚΑ για τον αριθμό επιβατικών οχημάτων κατά το έτος 2011 (Πηγές: ΥΠΕΚΑ, 2011· ΕΛΣΤΑΤ, 2018).

Υπό την παραπάνω οπτική, προκύπτει ότι για το έτος 2010 αντιστοιχούσαν περίπου 155 επιβατικά οχήματα σε κάθε χιλιόμετρο του εθνικού και επαρχιακού οδικού δικτύου της Ελλάδας. Ο αριθμός αυτός είναι ιδιαίτερα μεγάλος, εφόσον ληφθεί υπόψη ότι το μήκος 155 Ι.Χ. οχημάτων ξεπερνά τα 650m! Βέβαια, η θεώρηση αυτή αγνοεί την ύπαρξη των αστικών οδικών δικτύων, τα οποία επιφορτίζονται κατά κύριο λόγο το μεγαλύτερο ποσοστό της κυκλοφορίας των οχημάτων σε καθημερινή βάση, αλλά και τον κυκλοφοριακό φόρτο που προκύπτει από την κίνηση οχημάτων της αλλοδαπής. Παρόλα αυτά, ακόμη και σε αυτή την απλουστευμένη μορφή, η παραπάνω τιμή της γραμμικής πυκνότητας υποδεικνύει την τάξη μεγέθους της εν δυνάμει φόρτισης του διαπεριφερειακού δικτύου της χώρας.

Τέλος, σύμφωνα με την Ελληνική Στατιστική Αρχή (ΕΛΣΤΑΤ, 2008), από τα οχήματα που κυκλοφόρησαν για πρώτη φορά τη διετία 2006-2007, το 67% ήταν *επιβατικά αυτοκίνητα*, το 11,5% *φορτηγά*, το 0,25% *λεωφορεία* και το 21,10% *μοτοσυκλέτες*. Οι τιμές αυτές συνιστούν μία ένδειξη για την κατανομή του τύπου των οχημάτων που πραγματοποιούν συστηματική χρήση του ελληνικού διαπεριφερειακού δικτύου, η οποία παρέχει περαιτέρω πληροφορία για τη δομή και τη λειτουργία του.

*3.2. Το σιδηροδρομικό δίκτυο*
3.2.1. Ιστορικό πλαίσιο
Σύμφωνα με τον Οργανισμό Σιδηροδρόμων Ελλάδος - ΟΣΕ (2015α), η ιστορία των ελληνικών σιδηροδρόμων ξεκινά το έτος 1869, όπου πραγματοποιούνται τα εγκαίνια του συρμού για την έναρξη των δρομολογίων Θησείου – Πειραιά. Δεκατέσσερα χρόνια μετά, το 1882, συστήνεται η ανώνυμη μετοχική εταιρεία υπό την επωνυμία *Σιδηρόδρομοι Αθηνών – Πειραιώς – Πελοποννήσου* (ΣΠΑΠ), με έδρα την Αθήνα.

Το 1884 εγκαινιάζεται ο *Θεσσαλικός Σιδηρόδρομος* με πρώτο δρομολόγιο τη διαδρομή Βόλος-Λάρισα. Επτά χρόνια μετά, το 1890, ιδρύεται η εταιρεία *Σιδηροδρόμου Βορειοδυτικής Ελλάδος* (ΣΒΔΕ), η οποία εγκαινιάζει τη λειτουργία της γραμμής Μεσολογγίου – Αγρινίου.

Μέχρι τις αρχές του 20$^{ου}$ αιώνα οι σιδηροδρομικές μεταφορές στην Ελλάδα εκτελούνταν υπό την αιγίδα περιφερειακών οργανισμών, δίχως την εποπτεία ενός κεντρικού κρατικού φορέα. Για την κάλυψη αυτής της ανάγκης ιδρύονται το 1920 οι *Σιδηρόδρομοι Ελληνικού Κράτους* (ΣΕΚ), που αποτελούν νομικό πρόσωπο δημοσίου δικαίου με στόχο την ενοποίηση των μέχρι τότε ενεργών περιφερειακών σιδηροδρόμων του Ελληνικού Κράτους και την ανασυγκρότησή τους. Η διαδικασία αυτή ολοκληρώνεται το 1965, ενώ έξι χρόνια μετά, το 1970, ιδρύεται ο *Οργανισμός Σιδηροδρόμων Ελλάδος* (ΟΣΕ), στη σημερινή του μορφή, με σκοπό την οργάνωση, εκμετάλλευση και ανάπτυξη των σιδηροδρομικών μεταφορών της χώρας.

Η σύγχρονη ιστορία του ΟΣΕ έχει να επιδείξει το 1997 τη λειτουργία της πρώτης ηλεκτροκίνητης σιδηροδρομικής γραμμής Θεσσαλονίκης – Ειδομένης, ενώ έντεκα χρόνια μετά, το 2007, πραγματοποιήθηκε ο οργανωτικός επιμερισμός της Υποδομής από την Εκμετάλλευση των Σιδηροδρόμων, με τη δημιουργία των εταιρειών ΕΔΙΣΥ ΑΕ και ΤΡΑΙΝΟΣΕ ΑΕ, η δεύτερη εκ των οποίων ανεξαρτητοποιήθηκε από τον Όμιλο το 2008, υπαγόμενη απευθείας στο Ελληνικό Δημόσιο.

Στις αρχές της τρέχουσας δεκαετίας ολοκληρώθηκε η απορρόφηση της πρώην θυγατρικής εταιρείας ΕΔΙΣΥ ΑΕ από τον ΟΣΕ ΑΕ, γεγονός που σήμανε την άμεση υλοποίηση του νέου θεσμικού πλαισίου για την αναδιάρθρωση και τον εκσυγχρονισμό του ΟΣΕ. Το 2011 πραγματοποιήθηκε εξορθολογισμός των δαπανών, του ανθρωπίνου δυναμικού και της οργανωτικής δομής του Οργανισμού, επιφέροντας για πρώτη χρονιά θετικά λειτουργικά αποτελέσματα, σε σχέση με τα προηγούμενα έτη. Το 2012 το λειτουργικό αποτέλεσμα του ΟΣΕ προκύπτει επίσης θετικό για δεύτερη συνεχή χρονιά και εφαρμόστηκε η νέα οργανωτική δομή που ρύθμιζε την απόσχιση του κλάδου συντήρησης τροχαίου υλικού και την απορρόφησή του από τη νεοσύστατη *Ελληνική Εταιρεία Συντήρησης Σιδηροδρομικού Τροχαίου Υλικού* - ΕΕΣΣΤΥ ΑΕ, τη μεταβίβαση του τροχαίου υλικού του ΟΣΕ στο Δημόσιο ή σε φορέα του Δημοσίου και τη μεταβίβαση των μετοχών της εταιρείας ΓΑΙΑΟΣΕ ΑΕ στο Ελληνικό Δημόσιο.

3.2.2. Τεχνικά στοιχεία
Τα δίκτυα σιδηροδρομικών μεταφορών ανήκουν στην οικογένεια των *δικτύων υποδομών* (*infrastructure networks*) (Barthelemy, 2011), κατά αντιστοιχία με την προαναφερόμενη περίπτωση των δικτύων οδικών μεταφορών. Σύμφωνα με τον Οργανισμό Σιδηροδρόμων Ελλάδος Α.Ε. (ΟΣΕ, 2015β) ως σιδηροδρομικό δίκτυο (Σχήμα 7) ορίζεται «*το σύνολο της σιδηροδρομικής υποδομής που διαχειρίζεται η υπεύθυνη εταιρία για την εγκατάσταση και τη συντήρησή της*». Το σιδηροδρομικό δίκτυο της Ελλάδας εξυπηρετεί 28 πρωτεύουσες ελληνικών καποδιστριακών νομών.

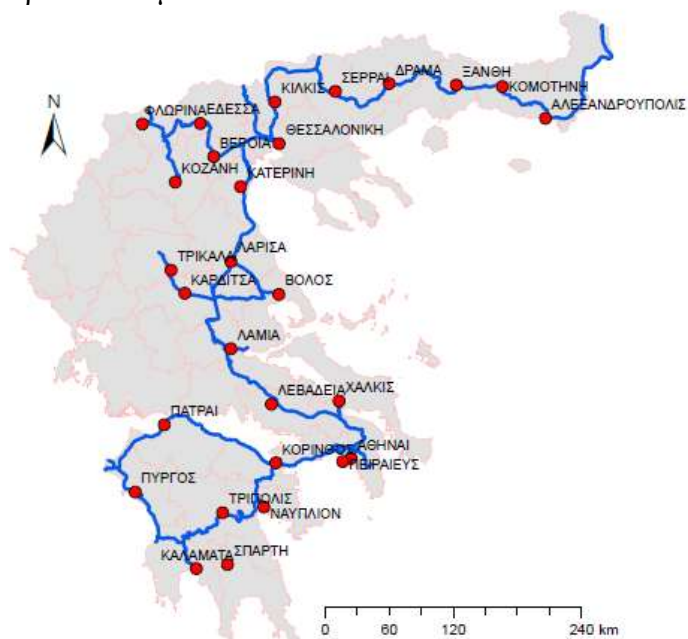

**Σχήμα 7.** Το δίκτυο σιδηροδρομικών μεταφορών της Ελλάδας (πηγή: geodata.gov.gr, 2010; ίδια επεξεργασία).

Στη Δήλωση Δικτύου του Οργανισμού, ορίζεται η έννοια της διαδρομής ως «*η χωρητικότητα της υποδομής που απαιτείται για να κινηθεί ένας συρμός μεταξύ δύο τόπων σε δεδομένο χρονικό διάστημα*», η οποία αντιστοιχεί στην έννοια του μονοπατιού (path) (Barthelemy, 2011) στην ορολογία της σύνθετης ανάλυσης δικτύων. Περαιτέρω, η έννοια του συγκοινωνιακού κόμβου ορίζεται ως «συγκεκριμένη γεωγραφική θέση/σταθμός του δικτύου που χρησιμεύει για την εξασφάλιση της κυκλοφορίας ή/και για την εξυπηρέτηση πελατών (επιβατών ή/και εμπορευμάτων)».

Σύμφωνα με την έκθεση πεπραγμένων του Οργανισμού για το έτος 2013 (ΟΣΕ, 2013), το σιδηροδρομικό δίκτυο του ΟΣΕ έχει συνολικό μήκος 2.773km, εκ των οποίων 2.265km αντιστοιχούν σε ενεργές γραμμές, 321km σε δίκτυο με αναστολή λειτουργίας και 187km σε καταργημένο δίκτυο, όπως φαίνεται στον Πίνακα 7.

**Πίνακας 7**
Το Δίκτυο του ΟΣΕ για το έτος 2013

| | Μήκος (m) | | | |
|---|---|---|---|---|
| | **Ενεργό Δίκτυο** | | | |
| **Είδος γραμμής** | *σε λειτουργία* | *σε προσωρινή αναστολή* | **Κατηργημένο Δίκτυο** | **Σύνολο** |
| Κανονικού πλάτους μονή ηλεκτροδοτούμενη | 82 | 0 | 0 | 82 |
| Κανονικού πλάτους μονή μη ηλεκτροδοτούμενη | 1.200 | 34 | 74 | 1.307 |
| Κανονικού πλάτους διπλή ηλεκτροδοτούμενη | 355 | 0 | 0 | 355 |
| Κανονικού πλάτους διπλή μη ηλεκτροδοτούμενη | 168 | 0 | 0 | 168 |
| Μετρικού πλάτους | 393 | 275 | 113 | 782 |
| Συνδυασμένου εύρους | 29 | 0 | 0 | 29 |
| Πλάτους 0,75m | 22 | 0 | 0 | 22 |
| Πλάτους 0,60m | 16 | 12 | 0 | 28 |
| **Σύνολο** | **2.265** | **321** | **187** | **2.773** |

(πηγή: ΟΣΕ, 2013)

Στον Πίνακα 8 παρουσιάζονται τα τμήματα του ενεργού δικτύου ΟΣΕ, όπως αυτά ίσχυσαν το έτος 2014.

**Πίνακας 8**
Τα τμήματα του ενεργού δικτύου ΟΣΕ για το έτος 2015

| ΑΞΟΝΕΣ | ΤΜΗΜΑΤΑ |
|---|---|
| ΚΥΡΙΟΣ ΑΞΟΝΑΣ (κανονικού εύρους) | Πειραιάς (Παλαιός σταθμός ΣΠΑΠ)-ΑΙΡ-Αθήνα-Οινόη-Λειανοκλάδι-Παλαιοφάρσαλος-Λάρισα-Πλατύ-Θεσσαλονίκη-Ειδομένη (Συνοριακός Σταθμός) |
| Διακλαδώσεις κύριου άξονα | Οινόη-Χαλκίδα |
| | Λιανοκλάδι – Λαμία – Στυλίδα |
| | Παλαιοφάρσαλος– Καλαμπάκα |
| | Λάρισα-Βόλος |
| ΑΞΟΝΑΣ ΠΡΟΑΣΤΙΑΚΟΥ ΑΘΗΝΩΝ | Αεροδρόμιο (Ελ. Βενιζέλος) – Μεταμόρφωση – ΣΚΑ – Λιόσια – Κόρινθος – Κιάτο |
| Διακλαδώσεις | Ν. Ικόνιο – ΧΘ 25+286 |
| | Αθήνα – Λιόσια |
| | Αθήνα – Μεταμόρφωση |
| ΔΥΤΙΚΗ ΜΑΚΕΔΟΝΙΑ (κανονικού εύρους) | (Θεσσαλονίκη-) Πλατύ – Έδεσσα – Αμύνταιο |
| Διακλαδώσεις | Αμύνταιο-Φλώρινα |
| | Αμύνταιο-ΧΣ 32+500 ΑμΚΖ |
| ΑΝΑΤΟΛΙΚΗ ΜΑΚΕΔΟΝΙΑ | Θεσσαλονίκη – Στρυμόνας – Αλεξανδρούπολη – Πύθιο – Δίκαια – Ορμένιο –Χ.Σ. 32+900 μΕΚ (Σύνορα) |

| ΑΞΟΝΕΣ | ΤΜΗΜΑΤΑ |
|---|---|
| (κανονικού εύρους) | |
| Διακλαδώσεις | Στρυμόνας-Προμαχώνας (Συνοριακός Σταθμός) |
| ΠΕΛΟΠΟΝΝΗΣΟΣ (μετρικού εύρους) | Ψαθόπυργος – Ρίο – Πάτρα – Αγ. Ανδρέας |
| | Αγ. Ανδρέας – Πύργος – Καλόνερο – Ζευγολατιό – Καλαμάτας (πλην Τακτικών Επιβατικών) |
| | Κόρινθος (Παλαιός Σταθμός)-Άργος-Τρίπολη (πλην Τακτικών Επιβατικών) |
| | Κόρινθος (Παλαιός Σταθμός)-Κόρινθος (Νέος Σταθμός) (πλην Τακτικών Επιβατικών) |
| | Αγ. Ανάργυροι-Ελευσίνα (συνδυασμένου εύρους) |
| Διακλαδώσεις | Διακοφτό-Καλάβρυτα |
| | Πύργος-Ολυμπία |
| | Πύργος - Κατάκολο |
| | Άργος-Ναύπλιο (πλην Τακτικών Επιβατικών) |
| ΓΡΑΜΜΗ ΠΗΛΙΟΥ (600 mm) | Άνω Λεχώνια - Μηλιές |

(πηγή: ΟΣΕ, 2015β)

Η ανώτατη ταχύτητα κίνησης εντός του δικτύου ανέρχεται στα 160km/*h*, η οποία εφαρμόζεται στο 18% του σιδηροδρομικού δικτύου, επίσης στο 18% εφαρμόζεται ταχύτητα μικρότερη των 79km/*h*, στο 40% εφαρμόζεται ταχύτητα 80-119km/*h* και στο υπόλοιπο 24% εφαρμόζεται ταχύτητα 120-159km/*h* (ΟΣΕ, 2013). Σύμφωνα με τη Δήλωση Δικτύου του Οργανισμού (ΟΣΕ, 2015β), στο σύνολο των εξυπηρετούμενων κόμβων του σιδηροδρομικού δικτύου της χώρας περιλαμβάνονται συνοριακοί, τερματικοί, επιβατικοί και εμπορευματικοί σταθμοί. Οι συνοριακοί σταθμοί αποτελούν τα σημεία σύνδεσης του δικτύου με τα αντίστοιχα σιδηροδρομικά δίκτυα των γειτονικών χωρών, οι τερματικοί σταθμοί τους κόμβους τερματισμού της σιδηροδρομικής υποδομής και οι επιβατικοί και εμπορευματικοί σταθμοί τους κόμβους επιβίβασης επιβατών και φόρτωσης εμπορευμάτων αντίστοιχα.

Η λειτουργία του σιδηροδρομικού δικτύου της χώρας υπόκειται σε κριτήρια προτεραιότητας για συγκεκριμένες υπηρεσίες (ΟΣΕ, 2015β), με σκοπό την εξασφάλιση της παροχής κατάλληλων υπηρεσιών μεταφορών, λαμβάνοντας υπόψη την κοινωνική σημασία της κατά προτεραιότητα ορισμένης υπηρεσίας σε σχέση με τις αντίστοιχες αποκλειόμενες. Βασικά κριτήρια προτεραιότητας αποτελούν η εξυπηρέτηση πρωτίστως των γραμμών intercity, έπειτα των προαστιακών, στη συνέχεια των κανονικών επιβατηγών και τέλος των εμπορευματικών.

Το νομικό πλαίσιο για την πρόσβαση στην σιδηροδρομική υποδομή και τις συναφείς υπηρεσίες, περιγράφεται στο Κεφάλαιο V του Προεδρικού Διατάγματος με αριθμό 41/2005 «Εναρμόνιση της ελληνικής νομοθεσίας με τις οδηγίες 91/440/ΕΟΚ και 95/18/ΕΟΚ όπως τροποποιήθηκαν με τις οδηγίες 2001/12/ΕΚ και 2001/13/ΕΚ, αντιστοίχως και της οδηγίας 2001/14/ΕΚ για την ανάπτυξη των Κοινοτικών σιδηροδρόμων, τις άδειες σε σιδηροδρομικές επιχειρήσεις, την κατανομή της χωρητικότητας των σιδηροδρομικών υποδομών και τις χρεώσεις για τη χρήση σιδηροδρομικής υποδομής και την πιστοποίηση ασφάλειας, και κατάργηση των ΠΔ 324/1996, 76/1998 και 180/1998», όπως ισχύει σήμερα (ΟΣΕ, 2015β).

Τέλος το τροχαίο υλικό του σιδηροδρομικού δικτύου διακρίνεται, ως προς το μηχανολογικό του ρόλο, σε *έλκον* (άμαξες) και *ρυμουλκούμενο* (βαγόνια). Το ρυμουλκούμενο τροχαίο υλικό διακρίνεται με τη σειρά του, ως προς τον οικονομικό του ρόλο, σε *επιβατηγό* και *εμπορευματικό* (ΟΣΕ, 2013), όπως φαίνεται αναλυτικά στον Πίνακα 9.

**Πίνακας 9**
Το τροχαίο υλικό του ΟΣΕ

| Έλκον τροχαίο υλικό | | |
|---|---|---|
| Συνολικός αριθμός αμαξών | 308 | |
| Δηζελάμαξες | | 142 |
| Ηλεκτράμαξες | | 30 |
| Αυτοκινητάμαξες (Diesel) (DMU) | | 109 |
| Ατμάμαξες | | 7 |
| Ηλεκτροκίνητες Αυτοκινητάμαξες (EMU) | | 20 |
| **Ρυμουλκούμενο τροχαίο υλικό** | | |
| *Εμπορευματικό τροχαίο υλικό* | | |
| Συνολικός αριθμός βαγονιών | 3.184 | |
| Κλειστά | | 962 |
| Επίπεδα | | 752 |
| Άλλου τύπου | | 1.470 |
| *Επιβατηγό τροχαίο υλικό* | | |
| Συνολικός αριθμός βαγονιών | 724 | |
| Επιβατάμαξες | | 301 |
| Κλιματιζόμενα | | 253 |
| Εστιατόρια | | 28 |
| Κλινοθέσια | | 15 |
| Κλινάμαξες | | 31 |
| Κινητήρια και Ρυμουλκούμενα οχήματα αυτοκινηταμαξών | 423 | |
| Κλιματιζόμενα | | 344 |
| Λοιπά | | 79 |

(Πηγή: ΟΣΕ, 2013)

*3.3. Το δίκτυο ακτοπλοϊκών μεταφορών*
3.3.1. Εννοιολογικό και ιστορικό πλαίσιο
Ετυμολογικά (Μανδάλα, 1988), η λέξη ακτοπλοΐα προέρχεται από τις σύνθετες λέξεις *ακτή* (=παραλία, γιαλός) + *πλους* (=πλεύση) και αναφέρεται κυριολεκτικά στον πλου κοντά στην ακτή και γενικότερα στη θαλάσσια συγκοινωνία ανάμεσα σε λιμένες εντός της ίδιας χώρας. Η ακτοπλοΐα αποτέλεσε πιθανότατα, μαζί με την πεζοπορία, τον αρχαιότερο τρόπο συγκοινωνίας σε παραθαλάσσιες περιοχές, καθόσον η κατασκευή πλωτών εξέδρων και πλοιαρίων δεν προϋπόθετε την εφεύρεση του τροχού. Η πλεύση στην ακτοπλοΐα πραγματοποιείται κατά βάση «εν όψει ακτών», δηλαδή η θέση του πλοίου (το γεωγραφικό του στίγμα) προσδιορίζεται κάθε φορά από την οπτική επαφή με τη γεωμορφολογία της ξηράς και με τη συμπληρωματική χρήση βοηθημάτων (όπως πυξίδα, ναυτικοί χάρτες, φάροι, κλπ). Η ακτοπλοΐα, ως είδος ναυσιπλοΐας, εμπεριέχεται σε όλα τα είδη των θαλασσίων μεταφορών, διότι η αρχή και το τέλος ενός ναυτιλιακού ταξιδιού αναφέρονται πάντοτε σε χερσαία τοποθεσία, δηλαδή στον τόπο προέλευσης και προορισμού.

Η Ελλάδα αποτελεί μια μεσογειακή χώρα με τεράστια παράδοση στις θαλάσσιες μεταφορές, γεγονός που οφείλεται τόσο στη γεωγραφική θέση και στη γεωμορφολογία της, όσο και στο ανήσυχο πνεύμα των Ελλήνων, που έρεπε από αρχαιοτάτων χρόνων προς την περιπέτεια και τη γνώση. Γεωγραφικά, βρίσκεται ανάμεσα σε δύο ηπείρους (Ευρώπη, Ασία) και τρεις θάλασσες (Μαύρη θάλασσα, Μεσόγειος, Αδριατική) και καλύπτεται από το Αιγαίο αρχιπέλαγος που διαβρέχει περισσότερες από 1.350 νησιά και βραχονησίδες, εκ των οποίων περισσότερα από 230 είναι κατοικημένα (Ψαραυτής, 2006; Tsiotas and Polyzos, 2014). Η γεωμορφολογία της χώρας ευνόησε προφανώς την ανάπτυξη ενός ισχυρού συστήματος ναυσιπλοΐας, με αποτέλεσμα σήμερα να περιλαμβάνεται στις ισχυρότερες ναυτιλιακές δυνάμεις παγκοσμίως. Ήδη, ακόμη και μέσα από τα έργα του Ομήρου αντλούνται πληροφορίες που στοιχειοθετούν την ισχύ της ελληνικής ναυσιπλοΐας εκείνης της εποχής. Ενδεχομένως η χώρα μας να αποτελεί τη γενέτειρα του πλοίου και της ναυπηγικής τέχνης στη συστηματική τους μορφή, υπόθεση η οποία ενισχύεται από αναφορές σχετικά με τη ναυτική ζωή και τη ναυπηγική τέχνη, όπως στα αρχαία κείμενα του Ομήρου (Οδύσσεια, Ε/253-260) και του Ξενοφώντος (Αθηναίων Πολιτεία Ι/1-2). Ακόμη και το γεγονός της επικράτησης στη διεθνή ορολογία λέξεων όπως *nautical* (ναυτικός), *nautilus* (ναυτίλος), *nave* (πλήμνη) κλπ., υποδηλώνει την εξέχουσα σημασία της ελληνικής ναυσιπλοΐας και τη σημαντική επιρροή της παγκοσμίως.

3.3.2. Τεχνικά στοιχεία

Σήμερα η ελληνόκτητη εμπορική ναυτιλία, η οποία απαρτίζεται από πλοία ελληνικής κατοχής που νηολογούνται είτε υπό ελληνική είτε υπό ξένη σημαία, αποτελεί την πρώτη ναυτιλιακή δύναμη παγκοσμίως, καταμετρώντας στόλο 3.480 πλοίων (βάρους μεγαλύτερου του ενός τερατόνου – 1t*tn*), όλων των τύπων, τα οποία διαθέτουν συνολική χωρητικότητα 98.195.100gtn (γιγατόνων). Ο αριθμός αυτός αποτελεί το 15,5% της ναυτιλιακής χωρητικότητας παγκοσμίως, αντικατοπτρίζοντας τη δυναμική της ελληνόκτητης ναυτιλίας στο παγκόσμιο στερέωμα (ΥΝΑ, 2015α).

Η ελληνική ναυτιλία απαρτίζεται από τα νηολογημένα υπό ελληνική σημαία πλοία και κατατάσσεται – εδώ και πολλές δεκαετίες – ανάμεσα στις κορυφαίες θέσεις των μεγαλύτερων ναυτιλιακών δυνάμεων παγκοσμίως, κατέχοντας το 5% περίπου της χωρητικότητας του νηολογημένου στόλου παγκοσμίως (ΥΝΑ, 2015α,β).

Ο υπό ελληνική σημαία στόλος απαριθμεί σήμερα 1.455 πλοία, συνολικής χωρητικότητας 32.048.052gtn, τοποθετώντας τη χώρα στην έβδομη θέση της παγκόσμιας κατάταξης, ως προς τη χωρητικότητα του ναυτιλιακού στόλου. Σε ευρωπαϊκό επίπεδο, ο ελληνικός εμπορικός στόλος καταλαμβάνει την πρώτη θέση και μερίδιο χωρητικότητας το 24% του αντίστοιχου ευρωπαϊκού και το 40% περίπου της συνολικής χωρητικότητας του εμπορικού στόλου της ΕΕ. Η ηλικία των πλοίων που εγγράφονται στα ελληνικά νηολόγια ανέρχεται σε 9,4 έτη κατά μέσο όρο, ενώ η μέση ηλικία των πλοίων που διαγράφονται ανέρχεται στα 20,7 έτη, γεγονός που υποδηλώνει μία τάση εκσυγχρονισμού του ελληνικού στόλου (ΥΝΑ, 2015α).

Περαιτέρω, η ελληνική ποντοπόρος εμπορική ναυτιλία αναπτύσσει δράση σε διεθνές επίπεδο, εξυπηρετώντας σε ποσοστό περισσότερο του 95% της χωρητικότητας του στόλου της μεταφορικές ανάγκες τρίτων χωρών, δράση η οποία είναι γνωστή διεθνώς με τον όρο *cross trade*. Τέλος, στα ελληνικά νηολογημένα πλοία αυτά απασχολείται σημαντικός αριθμός εργαζομένων που ξεπερνά τις 25.000 (ΥΝΑ, 2015β).

Η υποδομή του ακτοπλοϊκού δικτύου μεταφορών της χώρας είναι δαιδαλώδης, γεγονός που οφείλεται στην πλούσια θαλάσσια γεωμορφολογία της Ελλάδας, η οποία επιβάλει την εξυπηρέτηση πολυάριθμων προορισμών. Ειδικότερα, σε ετήσια βάση η λειτουργία του ακτοπλοϊκού δικτύου της χώρας εξειδικεύεται στην κάλυψη των αναγκών των κατοίκων των ελληνικών νησιών και της εμπορευματικής κίνησης, ενώ κατά την καλοκαιρινή περίοδο επιφορτίζεται και με τον εποχιακό τουρισμό, ο οποίος αυξάνει πολλαπλασιαστικά την ακτοπλοϊκή κίνηση και τις λειτουργικές ανάγκες των ακτοπλοϊκών υποδομών (Tsiotas and Polyzos, 2015a).

Σύμφωνα με την Εθνική Στατιστική Υπηρεσία Ελλάδος (ΕΣΥΕ, 2008) το σύνολο των ελληνικών λιμένων που κατέγραψαν εμπορική και εμπορευματική κίνηση για το έτος 2008 ήταν 229 λιμάνια, ενώ 116 από αυτά ανέπτυσσαν αποκλειστικά εμπορευματική δραστηριότητα (Σχήμα 8).

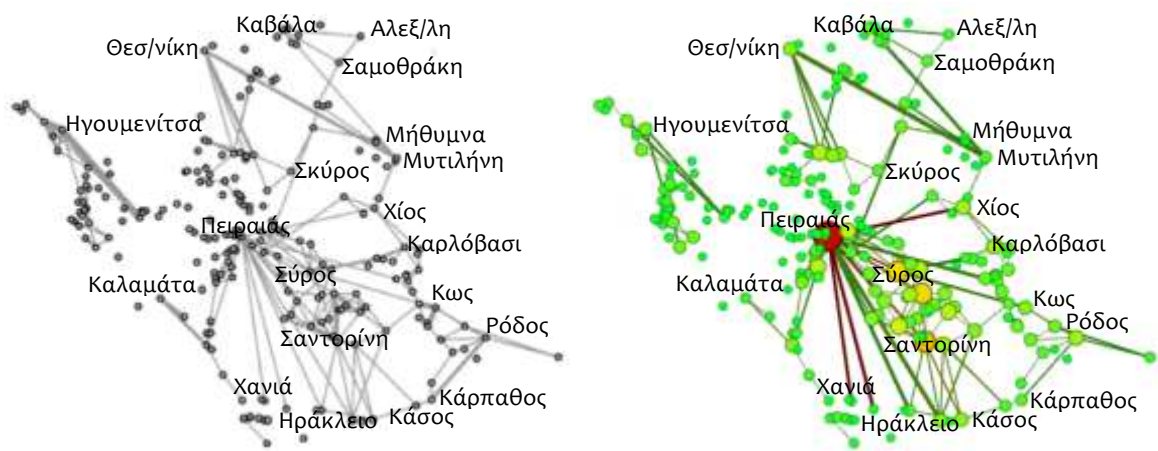

**Σχήμα 8.** (αρ.) Ο γράφος του ακτοπλοϊκού δικτύου επιβατηγών και εμπορικών μεταφορών της Ελλάδας (πηγή: Tsiotas and Polyzos, 2015) (δεξ.) Παρουσίαση της αξίας των λιμένων ανάλογα με τη συνδετικότητα τους (πηγή: ίδια επεξεργασία).

Είναι προφανές ότι η χαρτογράφηση και μελέτη του ακτοπλοϊκού δικτύου της χώρας αποτελεί ιδιαίτερα επίπονη διαδικασία, καθόσον ο καθορισμός των δρομολογίων επαφίεται στην ιδιωτική πρωτοβουλία των πολυάριθμων πλοιοκτητριών εταιρειών και ρυθμίζεται κατά βάση από τους κανόνες της ελεύθερης αγοράς. Η εποχιακή διαφοροποίηση της ακτοπλοϊκής κίνησης είναι ιδιαίτερα σημαντική, στερώντας πολλές γραμμές από το ελληνικό ακτοπλοϊκό δίκτυο μεταφορών, με αποτέλεσμα να ανακύπτουν θέματα διαχείρισης των άγονων γραμμών που επιδοτούνται κατά περίπτωση από το κράτος για τη διατήρηση των συγκοινωνιών και της επικοινωνίας με τις απομακρυσμένες και άγονες περιοχές (Tsiotas and Polyzos, 2015a).

Κομβικό ρόλο στη λειτουργία του ακτοπλοϊκού δικτύου της Ελλάδας διαδραματίζει το λιμάνι του Πειραιά, το οποίο διαθέτει σύγχρονες υποδομές και δυναμική να αναδειχθεί ως διεθνές ναυτιλιακό κέντρο. Ειδικότερα, στον Πειραιά φιλοξενούνται πάνω από 1.200 ναυτιλιακές εταιρείες της αλλοδαπής, οι οποίες διαχειρίζονται το σύνολο του ελληνόκτητου στόλου τους (που ξεπερνά τα 3200 ποντοπόρα πλοία με ελληνική σημαία), απασχολώντας στις δραστηριότητές τους περισσότερους από 12.300 εργαζομένους. Επιπρόσθετα στο λιμάνι του Πειραιά δραστηριοποιούνται πολυάριθμες επιχειρήσεις που σχετίζονται με τη ναυτιλία, καθώς και με παραναυτιλιακές δραστηριότητες (όπως ναυλωτές, ναυπηγεία, πρακτορεία, επιχειρήσεις τροφοδοσίας και καυσίμου, ναυτικά δικαστήρια, κλπ) (ΥΝΑ, 2015β).

Σύμφωνα με το Ν.2932/01 και τον ευρωπαϊκό Κανονισμό 3577/92/ΕΟΚ, οι Έλληνες και οι κοινοτικοί πλοιοκτήτες έχουν τη δυνατότητα να δρομολογήσουν ελεύθερα τα πλοία τους στις θαλάσσιες ενδομεταφορές, σε γραμμές της επιλογής τους, ανάλογα με την επιχειρηματική τους πρωτοβουλία. Η δρομολόγηση των πλοίων πραγματοποιείται – πλην εξαιρέσεων που εξετάζονται κατά περίπτωση – ετησίως, κατόπιν δηλώσεως που υποβάλλουν οι πλοιοκτήτριες εταιρείες εντός του πρώτου μήνα του έτους, στην οποία αναφέρονται οι επιθυμητές γραμμές που πρόκειται να εξυπηρετηθούν κατά την δρομολογιακή περίοδο από 1$^\text{η}$ Νοεμβρίου του ίδιου έτους μέχρι 31$^\text{η}$ Οκτωβρίου του επομένου (ΥΝΑ, 2015γ).

Στο παραπάνω πλαίσιο, ο κρατικός έλεγχος περιορίζεται στον έλεγχο ύπαρξης των απαιτούμενων κατά νόμο προϋποθέσεων δρομολόγησης και στην εξασφάλιση της προστασίας του δημοσίου συμφέροντος. Ειδικότερα, το αρμόδιο Υπουργείο (Ναυτιλίας και Αιγαίου) έχει την, κατ' εξαίρεση και στον αναγκαίο βαθμό, δυνατότητα παρέμβασης στις ελεύθερες δρομολογήσεις των πλοίων, στις περιπτώσεις κατά τις οποίες ανακύπτουν ζητήματα ασφάλειας της ναυσιπλοΐας στους λιμένες και τάξης στη χερσαία ζώνη, αλλά και στις οποίες παρακωλύεται η τακτική παροχή υπηρεσιών συγκεκριμένων γραμμών (ΥΝΑ, 2015γ).

Στις γραμμές που δεν επιλέγονται προς εξυπηρέτηση, με βάση τα επιχειρηματικά τους κριτήρια των πλοιοκτητών, εκδίδονται προσκλήσεις για τη σύναψη συμβάσεων ανάθεσης δημοσίων υπηρεσιών, διάρκειας μέχρι και 12 ετών. Η έκδοση των προσκλήσεων πραγματοποιείται με γνωμοδότηση του Συμβουλίου Ακτοπλοϊκών Συγκοινωνιών, η σύνθεση του οποίου αποτελείται από εκπροσώπους των εμπλεκομένων στην ακτοπλοΐα επαγγελματικών και κοινωνικών φορέων και της τοπικής και νομαρχιακής αυτοδιοίκησης. Με απόφαση του Υπουργού, έπειτα από γνωμοδότηση του Συμβουλίου, είναι δυνατόν να επιβάλλονται στους πλοιοκτήτες υποχρεώσεις δημόσιας υπηρεσίας αναφορικά με τους λιμένες, την τακτικότητα, τη συνέχεια, την ικανότητα παροχής μεταφορικών υπηρεσιών, το ναυλολόγιο και τη στελέχωση (ΥΝΑ, 2015γ).

Τέλος, θέματα σχετικά με το εργατικό προσωπικό των πλοίων άπτονται των διατάξεων της ελληνικής νομοθεσίας. Όσοι από τους ναυτικούς δεν κατέχουν την ελληνική ιθαγένεια οφείλουν να είναι πιστοποιημένοι για την ελληνομάθειά τους. Αρμόδια σε θέματα ανταγωνισμού στις θαλάσσιες ενδομεταφορές είναι η Επιτροπή Ανταγωνισμού (ΥΝΑ, 2015γ).

*3.4. Το εθνικό δίκτυο αεροπορικών μεταφορών*
3.4.1. Ιστορικό πλαίσιο
Σύμφωνα με την Fragoudaki (2000), οι επιχειρήσεις των εμπορικών αεροπορικών μεταφορών άρχισαν στην Ελλάδα το 1931 με τη δραστηριοποίηση ενός μοναδικού αερομεταφορέα, την *Ελληνική Εταιρεία Εναερίων Συγκοινωνιών* (ΕΕΕΣ) που επιχειρούσε με στόλο τεσσάρων Junkers G24, δώδεκα θέσεων το καθένα. Έχοντας ως πρώτο επιβάτη τον πρωθυπουργό της Ελλάδας, Ελευθέριο Βενιζέλο, η ΕΕΕΣ δραστηριοποιήθηκε για εννέα συναπτά έτη, μέχρι την έναρξη του Β΄ Παγκοσμίου Πολέμου, εξυπηρετώντας συνολικά 6.690 εγχώριους επιβάτες.

Το 1946, μετά τη λήξη του Β΄ παγκοσμίου πολέμου, οι αεροπορικές επιχειρήσεις συνεχίστηκαν, αυτή τη φορά από μια διαφορετική εταιρία, ενώ το 1947 παραχωρήθηκαν τρεις επιπλέον άδειες αερομεταφορέων. Το 1951 η ελληνική κυβέρνηση αποφάσισε να συγχωνεύσει τις τρεις αυτές εταιρείες για τη δημιουργία ενός εθνικού αερομεταφορέα, ενώ το 1955, προσπαθώντας να ξεπεράσει τα οικονομικά προβλήματα που ανέκυψαν, προέβη στην παραχώρηση του αποκλειστικού δικαιώματος διενέργειας εγχώριων αερομεταφορών στο μεγιστάνα Αριστοτέλη Ωνάση, με την πώληση του αποκλειστικού δικαιώματος διενέργειας εγχώριων αερομεταφορών, δημιουργώντας πλήρες μονοπώλιο στη διεξαγωγή των ελληνικών αεροπορικών μεταφορών. Η εν λόγω παραχώρηση περιλάμβανε το σύνολο του πτητικού έργου, συμπεριλαμβανομένων των πτήσεων ελικοπτέρων, της επισκευή και συντήρηση αεροσκαφών και τις υπηρεσίες επίγειας εξυπηρέτησης Το γεγονός αυτό σηματοδότησε την απαρχή της ιστορικής διαδρομής της *Ολυμπιακής Αεροπορίας* (Tsiotas and Polyzos, 2015b).

Το 1975 η ελληνική κυβέρνηση επανέκτησε την πλήρη ιδιοκτησία της Ολυμπιακής Αεροπορίας από τον Ωνάση, συνεχίζοντας το καθιερωμένο μονοπώλιο των εθνικών αερομεταφορών. Μία από τις βασικές συνέπειες του μονοπωλίου αυτού υπήρξε ο περιορισμός της λειτουργίας των *εκμισθωμένων αεροπορικών πτήσεων* (*charter*), οι οποίες - μεταξύ άλλων - στερούνταν του δικαιώματος παροχής υπηρεσιών με αφετηρία την Ελλάδα και υποχρεούνταν να εξυπηρετήσουν τους εισερχόμενους στη χώρα τουρίστες, για τη μετάβαση στον προορισμό τους, πραγματοποιώντας ενδιάμεση στάση στην Αθήνα και μετεπιβίβαση σε πτήση εσωτερικού της Ολυμπιακής. Εξ αιτίας των περιορισμών αυτών, ο αερολιμένας των Αθηνών υποδεχόταν πάνω από το 70% των διεθνών αφίξεων εκείνη την περίοδο (ETEM, 2010; Tsiotas and Polyzos, 2015b).

Από τη δεκαετία του '80 η ελληνική πολιτεία άρχισε σταδιακά να αναγνωρίζει τη σημασία των εταιρειών εκμίσθωσης πτήσεων στην τουριστική και γενικότερα στην οικονομική ανάπτυξη της χώρας, χαλαρώνοντας τους ισχύοντες περιορισμούς για τη λειτουργία τους. Ειδικότερα, το 1982 επιτράπηκε στα αεροδρόμια της Λήμνου, της Λέσβου, της Μυκόνου και της Ζακύνθου η υποδοχή των πρώτων διεθνών πτήσεων charter. Επιπρόσθετα, στο πλαίσιο εναρμόνισης της εθνικής νομοθεσίας με τις σχετικές οδηγίες και τους κανονισμούς της ΕΕ, ψηφίστηκε το 1992 το ΠΔ.276/91, το οποίο σηματοδότησε την απαρχή της απελευθέρωσης των εθνικών αερομεταφορών, επιτρέποντας την πραγματοποίηση πτήσεων με αφετηρία την Ελλάδα και προορισμούς τις χώρες μέλη της ΕΕ, αλλά και την εκτέλεση μη προγραμματισμένων πτήσεων εσωτερικού από ελληνικές ιδιωτικές αεροπορικές εταιρείες. Η οριστική απελευθέρωση των αερομεταφορών συντελέστηκε όμως λίγο αργότερα, όταν η κυβέρνηση απέσυρε το μονοπώλιο της Ολυμπιακής Αεροπορίας στις προγραμματισμένες πτήσεις εσωτερικού, αρχικά το 1996 για τις διαδρομές εντός της ηπειρωτικής χώρας και στη συνέχεια το 1998 για τα ελληνικά νησιά (ETEM, 2010; Tsiotas and Polyzos, 2015b).

Η Ολυμπιακή Αεροπορία λειτούργησε κάτω από μια κατάσταση μονοπωλίου για σχεδόν 35 έτη, έχοντας έδρα την Αθήνα. Στο διάστημα που βρίσκονταν υπό την κρατική ιδιοκτησία ανέπτυξε δύο θυγατρικές εταιρείες, εκ των οποίων η πρώτη λειτούργησε ως ο βασικός αερομεταφορέας, ενώ η δεύτερη ως *εταιρεία χαμηλού κόστους* (*Low Cost Carrier* - LCC). Κατά τη διάρκεια της περιόδου 1976-1999 η εταιρεία υποβλήθηκε σε 30 κορυφαίες διοικητικές αλλαγές, με μέση διάρκεια ζωής 9 μηνών έκαστη, ενώ ο στρατηγικός της προσανατολισμός καθοδηγούνταν σε μεγάλο βαθμό από κοινωνικά και πολιτικά κίνητρα. Ενδεικτικές δράσεις κυβερνητικών παρεμβάσεων στην άσκηση της διοίκησης της Ολυμπιακής υπήρξαν η πραγματοποίηση πτήσεων τακτικού χαρακτήρα σε απομακρυσμένες και απομονωμένες περιοχές, όπως μικρά νησιά και περιοχές με ανεπαρκή εναλλακτική μεταφορική κάλυψη, αλλά και η προσφορά τιμολογιακών προνομίων, όπως εκπτώσεις εισιτηρίου, σε επιλεγμένες κοινωνικές ομάδες (Fragoudaki, 2000; Tsiotas and Polyzos, 2015b).

Η αεροπορική εταιρία *Αερογραμμές Αιγίου* (*Aegean Airlines*) ιδρύθηκε το 1987 υπό την ονομασία *Αεροπορία Αιγίου* (*Aegean Aviation*), έχοντας έδρα στην Αθήνα. Το 1992 αποτέλεσε τον πρώτο ιδιωτικό ελληνικό μεταφορέα που απέκτησε άδεια χειριστή και το 1994 ξεκίνησε τη διεξαγωγή πτήσεων εξεχουσών προσωπικοτήτων (VIP) σε όλο τον κόσμο, με στόλο από ιδιόκτητα μικρά αεροσκάφη. Το 1999 εγκαινιάστηκε η επίσημη ίδρυσή της με την πραγματοποίηση των πρώτων πτήσεων από την Αθήνα προς τη Θεσσαλονίκη και το Ηράκλειο, χρησιμοποιώντας 2 ιδιόκτητα μεγάλα αεροσκάφη. Το 2001 συγχωνεύθηκε με μια άλλη ιδιωτική εταιρία, τις *Αερογραμμές Κρόνος* (*Cronus Airlines*), αυξάνοντας το στόλο και το αεροπορικό δίκτυό της. Το 2004 η Aegean Airlines ομογενοποίησε το στόλο της διατηρώντας δύο τύπους αεροσκαφών και το 2005 προέβη σε περαιτέρω αύξηση στόλου, συνάπτοντας συνεργασία με μια από τις μεγαλύτερες γερμανικές αεροπορικές εταιρίες. Το 2007 οι Αερογραμμές Αιγίου εισήχθησαν στο χρηματιστήριο και το 2008 προέβησαν σε περαιτέρω αύξηση στόλου, λαμβάνοντας πιστοποίηση τυποποίησης (ISO). Το 2010 η Aegean έγινε μέλος ενός μεγάλου παγκόσμιου δικτύου αεροπορικής συμμαχίας (*Star Alliance*). Η συνεχής μεγέθυνση, σε συνδυασμό με τις προαναφερόμενες παθογένειες της ανταγωνιστικής σε αυτήν εταιρία *Ολυμπιακές Αερογραμμές* (Olympic Airlines - πρώην Ολυμπιακή Αεροπορία), αλλά και την καθιέρωση της οικονομικής κρίσης στην Ελλάδα, οδήγησε την Aegean Airlines, τον Οκτώβριο του 2013, να προβεί στην εξαγορά της Olympic Airlines, η οποία σηματοδότησε μία νέα εποχή στην εγχώρια αγορά αερομεταφορών (Tsiotas και Polyzos, 2015b).

Στην Ελλάδα, την περίοδο μετά το 2000 δραστηριοποιήθηκαν συνολικά 10 μικρές εταιρίες, οι οποίες διεξήγαγαν κυρίως έκτακτες μισθωμένες πτήσεις charter. Σήμερα, στην εγχώριο αγορά αερομεταφορών δραστηριοποιούνται τρεις LCC εταιρείες με υπηρεσίες μεταφοράς επιβατών, η *Sky Express*, η *Astra Airlines* και η *RyanAir*.

Η εταιρία *Sky Express* εδράζεται στην πόλη του Ηρακλείου, στην Κρήτη. Ιδρύθηκε στις αρχές του 2005 από μια κοινοπραξία δύο επενδυτών που υπήρξαν προηγούμενα μέλη της Ολυμπιακής Αεροπορίας. Τον Ιούλιο του 2005 η Sky Express άρχισε πτήσεις charter και εμπορευματικών μεταφορών, αεροταξί, ιατρικές υπηρεσίες έκτακτης ανάγκης, εκδρομές και περιηγήσεις. Σήμερα η εταιρεία επιχειρεί με κύρια έδρα το αεροδρόμιο του Ηρακλείου, έχοντας πάνω από 20 προορισμούς (Tsiotas και Polyzos, 2015b).

Η εταιρία *Astra Airlines* εδράζεται στην πόλη της Θεσσαλονίκης. Αποτελεί επιχείρηση LCC που υποστηρίζεται από ελληνικό ταξιδιωτικό πράκτορα και ξεκίνησε να επιχειρεί στην εγχώριο αεροπορική αγορά το 2008. Σήμερα επιχειρεί στην ελληνική αγορά με τακτικές ετήσιες πτήσεις, από Θεσσαλονίκη και Αθήνα, έχοντας 10 βασικούς προορισμούς (Tsiotas και Polyzos, 2015).

Τέλος η εταιρεία *RyanAir* εταιρεία LCC εταιρεία Ιρλανδικών συμφερόντων που επιχειρεί στην ευρωπαϊκή αεροπορική αγορά. Έχει έδρα την Ιρλανδία και επιχειρησιακή βάση το αεροδρόμιο του Δουβλίνου το *Stansted* του Λονδίνου. Η εταιρεία ιδρύθηκε το 1985, έχει αναπτύξει στόλο 300 αεροσκαφών και επιχειρεί σε περισσότερους από 110 προορισμούς στην Ευρώπη, αποτελώντας τη μεγαλύτερη ευρωπαϊκή αεροπορική εταιρεία χαμηλού κόστους. Το 2010 ξεκίνησε να επιχειρεί και από ελληνικά αεροδρόμια σε ευρωπαϊκούς προορισμούς, ενώ το καλοκαίρι του 2014 εγκαινίασε πτήσεις εσωτερικού από Αθήνα και Θεσσαλονίκη προς Χανιά και Ρόδο, οι οποίες προς το παρόν διατηρούν τον εποχικό τους χαρακτήρα.

3.4.2. Τεχνικά στοιχεία

Σήμερα υφίστανται στην Ελλάδα 45 αεροδρόμια, από τα οποία τα 39 είναι ενεργά, ενώ τα υπόλοιπα 6 βρίσκονται σε αναστολή λειτουργίας. Από τα 39 αεροδρόμια που βρίσκονται σε λειτουργία, τα 34 αποτελούν κρατική ιδιοκτησία και διοικούνται από την Υπηρεσία Πολιτικής Αεροπορίας (ΥΠΑ), τα 4 αποτελούν ιδιοκτησία φορέων τοπικής αυτοδιοίκησης, αλλά διοικούνται εξίσου όμως από την ΥΠΑ, ενώ ο αερολιμένας *Ελευθέριος Βενιζέλος* των Αθηνών, λειτουργεί υπό το καθεστώς σύμβασης παραχώρησης στην ιδιωτική εταιρεία *Hochtief* και πρόκειται να περιέλθει στην κρατική ιδιοκτησία, μετά τη λήξη του συμβατικού χρόνου εκμετάλλευσης από την εταιρεία. Επιπλέον, 28 αεροδρόμια βρίσκονται σε νησιωτικές περιοχές, ενώ μόλις 11 εντός της ηπειρωτικής Ελλάδας, γεγονός που υποδηλώνει αφενός την ανταγωνιστικότητα του μέσου των εναερίων μεταφορών σε σχέση με αυτού των θαλασσίων μεταφορών και αφετέρου σκιαγραφεί τον αυξημένο όγκο των τουριστικών αναγκών της χώρας στα νησιά. Ειδικότερα, το 73% του συνόλου των αλλοδαπών τουριστών που επισκέπτονται τη χώρα αφικνούνται αεροπορικώς, ενώ το 60% των διαθέσιμων τουριστικών καταλυμάτων βρίσκεται στη νησιωτική επικράτεια (ΕΤΕΜ, 2010).

Ως προς τη γεωγραφία των πτήσεων, 15 από τα 39 ενεργά αεροδρόμια είναι χαρακτηρισμένα ως διεθνή, 11 εσωτερικού, ενώ 13 αποτελούν περιστασιακά οριζόμενα σημεία εισόδου εξόδου από τη χώρα. Στο Σχήμα 9 παρουσιάζεται η διαχρονική εξέλιξη του αριθμού των ενεργών αεροδρομίων στην Ελλάδα την περίοδο 1978-2008, στο οποίο παρατηρείται ο διπλασιασμός των νησιωτικών αεροδρομίων που εξυπηρετούν κίνηση εσωτερικού και ο τριπλασιασμός των νησιωτικών αεροδρομίων που εξυπηρετούν κίνηση εξωτερικού, ενώ η αντίστοιχη κατάσταση για τα ηπειρωτικά αεροδρόμια παρουσιάσει ανεπαίσθητη αύξηση. Το 2008 μετακινήθηκαν στην Ελλάδα 40.588.771 επιβάτες, εκ των οποίων το 31,23% αφορούσε πτήσεις εσωτερικού και το 68,77% πτήσεις εξωτερικού.

Περαιτέρω, από το συνολικό όγκο των πτήσεων εκείνης της χρονιάς, περίπου το 39,42% καλύφθηκε από πτήσεις *charter* (ΕΤΕΜ, 2010).

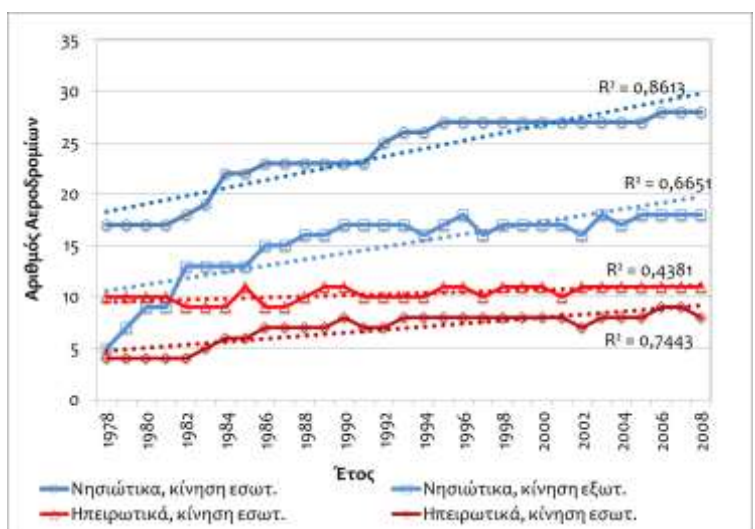

**Σχήμα 9.** Αριθμός ελληνικών αερολιμένων σε λειτουργία για την περίοδο 1978 – 2008 (πηγή: ΕΤΕΜ, 2010, ίδια επεξεργασία).

Σύμφωνα με κοινοτικό σύστημα ταξινόμησης αεροδρομίων (ΥΠΑ, 2013), ο αερολιμένας των Αθηνών αποτελεί αεροδρόμιο κατηγορίας Α (ετήσια κίνηση μεγαλύτερη των 10 εκατομμυρίων επιβατών), ο οποίος εξυπηρετεί το 39,79% του συνόλου των επιβατών της χώρας. Ο αερολιμένας του Ηρακλείου αποτελεί αεροδρόμιο κατηγορίας Β (με ετήσια κίνηση 5-10 εκατομμύρια επιβάτες), εξυπηρετώντας το 13,39% της συνολικής κίνησης. Περαιτέρω, 5 αεροδρόμια (Ρόδου, Θεσσαλονίκης, Κέρκυρας, Χανίων, Κω) κατηγορίας C (με ετήσια κίνηση 1-5 εκατομμύρια επιβάτες), τα οποία εξυπηρετούν το 34,84% της κίνησης. Τέλος, 32 αεροδρόμια κατηγορίας D (με ετήσια κίνηση μικρότερη του 1 εκατομμυρίου επιβατών), τα οποία εξυπηρετούν το 11,99% της κίνησης.

Αρκετά από τα αεροδρόμια της κατηγορίας D οφείλουν τη λειτουργία τους στις χορηγούμενες επιδοτήσεις μέσω του θεσμού των *Υποχρεώσεων Παροχής Δημόσιας Υπηρεσίας* (*Public Service Obligation* – PSO). Από την παραπάνω κατηγοριοποίηση προκύπτει ότι το ελληνικό αεροπορικό τοπίο χαρακτηρίζεται από υψηλή συγκέντρωση, καθόσον στα δύο μεγαλύτερα αεροδρόμια της χώρας, της Αθήνας και του Ηρακλείου, συντελείται περισσότερο από το 50% του συνόλου της επιβατικής κίνησης. Από την άλλη πλευρά, αριθμός μεγάλος αεροδρομίων λειτουργεί προς εξυπηρέτηση τοπικών αναγκών, δίχως να εξασφαλίζεται πάντοτε η οικονομική βιωσιμότητά τους (ΕΤΕΜ, 2010). Στον Πίνακα 10 παρουσιάζεται η συνολική αεροπορική κίνηση εσωτερικού και εξωτερικού των αερολιμένων στην Ελλάδα, για την περίοδο 1993-2013.

**Πίνακας 10**
Συνολική αεροπορική κίνηση εσωτερικού και εξωτερικού των αερολιμένων στην Ελλάδα, για την περίοδο 1993-2013

| Έτος | Αριθμός αεροσκαφών ΑΦ. & ΑΝ. | Αριθμός επιβατών | | Εμπορευματικό φορτίο (tn) | |
|---|---|---|---|---|---|
| | | Αφίξεις | Αναχωρήσεις | Φορτώσεις | Εκφορτώσεις. |
| 1993 | 276.897 | 11.800.468 | 11.851.830 | 67.430 | 58.019 |
| 1994 | 288.539 | 13.139.732 | 13.174.392 | 71.934 | 62.345 |
| 1995 | 292.365 | 13.024.974 | 13.064.271 | 75.164 | 64.884 |
| 1996 | 299.105 | 12.974.073 | 12.854.916 | 66.214 | 56.029 |
| 1997 | 332.491 | 14.276.024 | 13.794.811 | 93.278 | 65.271 |

| Έτος | Αριθμός αεροσκαφών ΑΦ. & ΑΝ. | Αριθμός επιβατών | | Εμπορευματικό φορτίο (tn) | |
|---|---|---|---|---|---|
| | | Αφίξεις | Αναχωρήσεις | Φορτώσεις | Εκφορτώσεις. |
| 1998 | 343.414 | 14.524.309 | 13.931.906 | 79.411 | 55.814 |
| 1999 | 396.624 | 16.458.544 | 16.346.191 | 82.677 | 52.553 |
| 2000 | 427.309 | 17.917.946 | 18.313.938 | 83.738 | 59.606 |
| 2001 | 396.192 | 17.600.688 | 17.957.823 | 81.432 | 55.102 |
| 2002 | 360.282 | 16.646.425 | 16.831.744 | 77.525 | 47.976 |
| 2003 | 395.773 | 16.971.795 | 17.053.431 | 75.780 | 50.732 |
| 2004 | 419.851 | 17.589.681 | 17.650.110 | 76.247 | 54.984 |
| 2005 | 403.163 | 18.130.097 | 18.316.739 | 73.387 | 51.987 |
| 2006 | 429.419 | 19.259.749 | 19.453.763 | 75.412 | 54.663 |
| 2007 | 455.199 | 20.529.838 | 20.784.541 | 74.541 | 55.489 |
| 2008 | 440.914 | 20.345.555 | 20.491.364 | 76.171 | 56.846 |
| 2009 | 457.706 | 19.746.105 | 19.899.381 | 63.139 | 52.088 |
| 2010 | 428.863 | 19.083.347 | 19.220.226 | 57.612 | 47.018 |
| 2011 | 410.226 | 19.322.847 | 19.508.474 | 48.398 | 42.373 |
| 2012 | 382.781 | 18.260.312 | 18.397.819 | 40.608 | 37.750 |
| 2013 | 375.362 | 19.173.277 | 19.284.064 | 38.753 | 37.722 |

(πηγή: ΥΠΑ, 2013)

Επιπρόσθετα, για την καλύτερη αξιολόγηση των αποτελεσμάτων του Πίνακα 10 κατασκευάστηκαν τα σχήματα 10, 11 και 12, τα οποία απεικονίζουν διαγραμματικά τη διαχρονική εξέλιξη των μεταβλητών του Πίνακα 10.

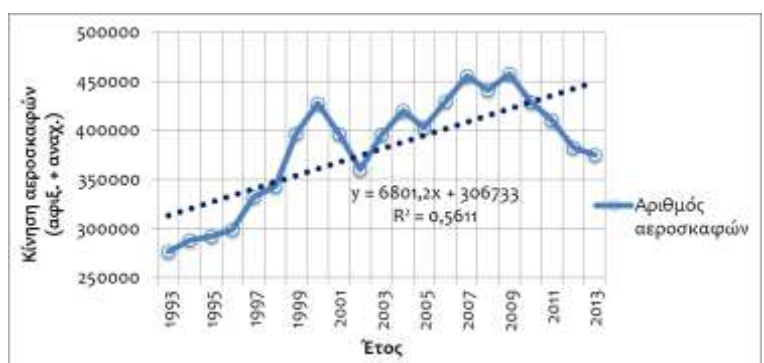

**Σχήμα 10.** Διαχρονική κίνηση αεροσκαφών (αφίξεις + αναχωρήσεις) στο σύνολο των ενεργών αεροδρομίων της Ελλάδας, κατά την περίοδο 1993-2013 (πηγή: ΥΠΑ, 2013).

Όπως προκύπτει από το Σχήμα 11, η αεροπορική κίνηση (κίνηση αεροσκαφών) στην Ελλάδα έχει σημειώσει διαχρονικά (από το 1993) αύξηση της τάξεως του 50%, αλλά είναι εμφανές ότι έχει επηρεαστεί από την εμφάνιση της οικονομικής κρίσης (έπειτα από το 2009), καταγράφοντας μείωση σε ποσοστό 18%. Επίσης, η αεροπορική επιβατική κίνηση έχει παρουσιάσει διαχρονικά αύξηση που έχει ξεπεράσει το 80%, αλλά ομοίως επηρεάστηκε από την εμφάνιση της κρίσης καταγράφοντας μείωση στο αντίστοιχο χρονικό διάστημα της τάξεως του 10%. Αντίθετα, η αεροπορική εμπορευματική κίνηση παρουσίασε διαχρονική μείωση της τάξεως του 60%, εμφανίζοντας κατά την περίοδο της κρίσης αυξημένους τους ρυθμούς μείωσης.

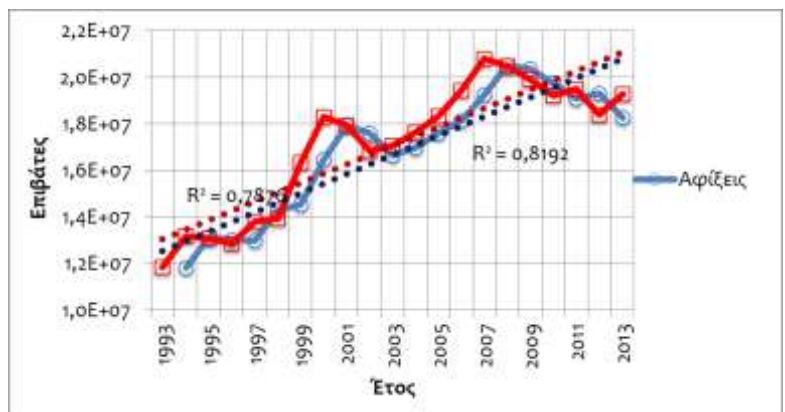

**Σχήμα 11.** Διαχρονική αεροπορική επιβατική κίνηση στο σύνολο των ενεργών αεροδρομίων της Ελλάδας, κατά την περίοδο 1993-2013 (πηγή: ΥΠΑ, 2013).

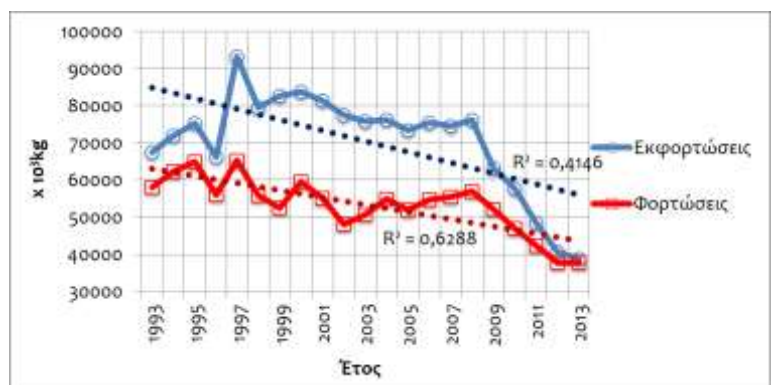

**Σχήμα 12.** Διαχρονική αεροπορική εμπορευματική κίνηση στο σύνολο των ενεργών αεροδρομίων της Ελλάδας, κατά την περίοδο 1993-2013 (πηγή: ΥΠΑ, 2013).

Σύμφωνα με τα εικονιζόμενα στοιχεία του Πίνακα 10 και των σχημάτων 3.11-3.13, η αεροπορική κίνηση στην Ελλάδα φαίνεται πως εμφανίζει εξειδίκευση στην εξυπηρέτηση επιβατών, καθόσον η μεταφορά εμπορευμάτων βαίνει διαχρονικά μειούμενη. Το γεγονός αυτό οφείλεται προφανώς στην ανταγωνιστικότητα του μέσου των θαλασσίων μεταφορών για τις εμπορικές μετακινήσεις στην Ελλάδα (Tsiotas and Polyzos, 2014), που υστερούν μεν σε ταχύτητα, αλλά υπερτερούν σε μεταφορικό κόστος, αλλά και στις σχετικά μικρές θαλάσσιες αποστάσεις που υφίστανται στις ελληνικές θάλασσες.

### 4. Εμπειρική Ανάλυση

Στην ενότητα αυτή πραγματοποιείται εμπειρική ανάλυση προκειμένου να διαπιστωθεί η ύπαρξη συναρτησιακής σχέσης μεταξύ των μεταφορικών υποδομών και του κοινωνικοοικονομικού περιβάλλοντός τους. Η ανάλυση στηρίζεται στην κατασκευή ενός οικονομετρικού υποδείγματος πολυμεταβλητής γραμμικής παλινδρόμησης (multivariate linear regression), το οποίο εκφράζει το επίπεδο ευημερίας των νομών (εξαρτημένη μεταβλητή $y$) ως συνάρτηση ($y=f(x_1,x_2,…,x_n)$) ενός συνόλου ανεξάρτητων μεταβλητών ($x_1,x_2,…,x_n$) που σχετίζονται με τις μεταφορικές υποδομές και το κοινωνικοοικονομικό πλαίσιο των νομών της χώρας. Οι μεταβλητές που χρησιμοποιούνται στην ανάλυση παρουσιάζονται στον Πίνακα 11 και ομαδοποιούνται ανάλογα με τη θεματική τους συνάφεια σε τρεις κατηγορίες, τις κοινωνικοοικονομικές, τις μεταβλητές μεταφορικών δικτύων και τις μικτές μεταβλητές που περιέχουν πληροφορία και από τις δύο προηγούμενες κατηγορίες.

**Πίνακας 11**
Μεταβλητές που συμμετέχουν στην εμπειρική ανάλυση

| Μεταβλητή | Συμβολισμός | Περιγραφή | Πηγή |
|---|---|---|---|
| | | **ΚΟΙΝΩΝΙΚΟΟΙΚΟΝΟΜΙΚΕΣ ΜΕΤΑΒΛΗΤΕΣ** | |
| $X_1$ | $T_{GDP}$ | Συμμετοχή του τουρισμού του νομού στο ΑΕΠ της χώρας. | Tsiotas and Polyzos (2015a) |
| $X_2$ | $B_{SEC}$ | Συμμετοχή του δευτερογενή τομέα του νομού στο ΑΕΠ της χώρας. | Πολύζος (2011) |
| $X_3$ | $C_{SEC}$ | Συμμετοχή του τριτογενή τομέα του νομού στο ΑΕΠ της χώρας. | Tsiotas and Polyzos (2015a) |
| $X_4$ | EDU | Επίπεδο εκπαίδευσης του πληθυσμού του νομού. | Πολύζος (2011) |
| Y | WELF | Επίπεδο ευημερίας του νομού. | Tsiotas and Polyzos (2015a) |
| | | **ΜΕΤΑΒΛΗΤΕΣ ΜΕΤΑΦΟΡΙΚΩΝ ΔΙΚΤΥΩΝ** | |
| $X_5$ | $ROAD_{DENS}$ | Πυκνότητα οδικού δικτύου του νομού. Υπολογίζεται από το λόγο του μήκους οδικού δικτύου προς την επιφάνεια του νομού. | Tsiotas (2017) |
| $X_6$ | $RAIL_{DENS}$ | Πυκνότητα σιδηροδρομικού δικτύου του νομού. Υπολογίζεται από το λόγο του μήκους σιδηροδρομικού δικτύου προς την επιφάνεια του νομού. | Tsiotas (2017) |
| $X_7$ | PORTS | Αριθμός λιμένων του νομού. | Tsiotas and Polyzos (2015a) |
| $X_8$ | AIRPORTS | Αριθμός αεροδρομίων του νομού. | Tsiotas and Polyzos (2015b) |
| | | **ΜΙΚΤΕΣ ΜΕΤΑΒΛΗΤΕΣ** **ΜΕΤΑΦΟΡΙΚΩΝ ΥΠΟΔΟΜΩΝ ΚΑΙ ΚΟΙΝΩΝ/ΚΟΥ ΠΕΡΙΒΑΛΛΟΝΤΟΣ** | |
| $X_9$ | TPP | Συνολικό πληθυσμιακό δυναμικό του νομού[α]. Υπολογίζεται από το λόγο του πληθυσμού προς την απόσταση μιας περιοχής προς την ίδια και τις γειτονικές της αγορές και εκφράζει το δυναμικό επιρροής των αγορών σε μια περιοχή. | Πολύζος (2011) |

α. Τιμές για τους 51 Καποδιστριακούς νομούς της χώρας

Ως εξαρτημένη μεταβλητή στο οικονομετρικό υπόδειγμα επιλέγεται το επίπεδο ευημερίας των νομών (WELF) γιατί θεωρείται πως αποτελεί αντιπροσωπευτικότερο δείκτη του επιπέδου ανάπτυξης μιας περιφέρειας σε σχέση πχ. με το ΑΕΠ που διατυπώνεται περισσότερο με όρους οικονομικής μεγέθυνσης (Πολύζος, 2011). Η μέθοδος που χρησιμοποιείται στο οικονομετρικό υπόδειγμα είναι η μέθοδος των αποβολών (backward elimination method - BEM) (Norusis, 2004; Walpole et al., 2012) κατά την οποία η διαδικασία ξεκινά με το σύνολο των διαθέσιμων μεταβλητών, παράγοντας μία αλληλουχία (σειρά) υποδειγμάτων με διαδοχική αποβολή της περισσότερο στατιστικά ασήμαντης μεταβλητής ($p$-value>0,1) σε κάθε κύκλο. Η μέθοδος τερματίζει στο σημείο όπου το σύνολο των μεταβλητών που έχουν απομείνει είναι στατιστικά σημαντικές για το υπόδειγμα. Για το σύνολο εξαρτημένων μεταβλητών του Πίνακα 11 $\mathbf{X}_9=\{X_1, X_2, ..., X_9\}$, η ακολουθία υποδειγμάτων the $(Y_k)_{k\geq 0}$ περιγράφεται από τις παρακάτω μαθηματικές σχέσεις:

$$(Y_k)_{k\in\{1,...,n\}\subseteq\mathbb{N}} \Big| Y_k = \sum_{i=1}^{9-k+1} b_i \cdot X_i + c_k \cdot \mathbf{1}$$

$$\begin{cases} \mathbf{X}_9 = \{X_1, X_2, ..., X_9\}, \\ X_i \in \mathbf{X}_{9-k+1}, \\ \mathbf{X}_{9-k} = X_{9-k+1} - \{\mathbf{x}_p\} \\ X_p \in \mathbf{X}_{9-k+1} : P[b(X_p)=0] = \max\{P[b_i=0] \geq 0,1\} \end{cases} \quad (1),$$

όπου $X_i$, $i=1,…,9$, είναι οποιαδήποτε μεταβλητή του Πίνακα 11, η οποία έχει 51 στοιχεία (κάθε ένα αντιστοιχεί σε μία τιμή της μεταβλητής για συγκεκριμένο νομό). Τα αποτελέσματα της πολυμεταβλητής ανάλυσης ΒΕΜ παρουσιάζονται στον Πίνακα 12. Η ικανότητα προσδιορισμού (βλ. υποπίνακα α) του υποδείγματος προκύπτει $R^2=0,638$ και εκφράζει ότι το 63,8% της μεταβλητότητας των ανεξάρτητων μεταβλητών περιγράφεται από το υπόδειγμα (δηλ. τη μεταβλητότητα της εξαρτημένης μεταβλητής).

**Πίνακας 12**
Αποτελέσματα του οικονομετρικού υποδείγματος για το επίπεδο ευημερίας των ελληνικών νομών

α. Σύνοψη

| Υπόδειγμα | R | $R^2$ | Προσαρμοσμένο $R^2$ | Τυπικό Σφάλμα της Εκτίμησης |
|---|---|---|---|---|
| Αποβολών (6ος κύκλος) | 0,799[f] | 0,638 | 0,607 | 12,59302 |

β. Συντελεστές[α,β]

| | | Μη τυποποιημένοι συντελεστές | | Τυποποιημένοι συντελεστές | | |
|---|---|---|---|---|---|---|
| Υπόδειγμα | | B | Std. Error | Beta | t | Σημαντικότητα |
| 6 | (Σταθερά) | 11,710 | 6,258 | | 1,871 | 0,068 |
| | $T_{GDP}$ | -0,018 | 0,007 | -0,537 | -2,743 | 0,009 |
| | EDU | 0,696 | 0,190 | 0,597 | 3,671 | 0,001 |
| | PORTS | 1,763 | 0,332 | 0,564 | 5,315 | 0,000 |
| | $ROAD_{DENS}$ | 0,051 | 0,026 | 0,274 | 1,965 | 0,056 |

α. Εξαρτημένη μεταβλητή: WELF
β. Ανεξάρτητες μεταβλητές: TPP, TGDP, EDU, $B_{SEC}$, $C_{SEC}$, PORTS, AIRPORTS, $RAIL_{DENS}$, $ROAD_{DENS}$

Όπως προκύπτει από τον Πίνακα 12(β), οι στατιστικά σημαντικές μεταβλητές που συμμετέχουν στο τελικό υπόδειγμα είναι το προϊόν τουρισμού ($T_{GDP}$), το επίπεδο εκπαίδευσης (EDU), ο αριθμός των λιμένων (PORTS) και η πυκνότητα του οδικού δικτύου ($ROAD_{DENS}$) των νομών. Η σχέση της μεταβλητής $T_{GDP}$ είναι αρνητική στο υπόδειγμα και εκφράζει ότι οι περιφέρειες με αυξημένο προϊόν στον τουριστικό τομέα τείνουν να εμφανίζουν μικρότερο επίπεδο ευημερίας. Οι υπόλοιπες μεταβλητές έχουν θετική συνεισφορά και εκφράζουν ότι οι περιφέρειες με υψηλό επίπεδο ευημερίας τείνουν να έχουν υψηλότερο επίπεδο εκπαίδευσης του πληθυσμού, μεγαλύτερο αριθμό λιμένων και πυκνότερο οδικό δίκτυο. Συνολικά, παρατηρείται ότι οι δύο από τις τέσσερις μεταβλητές που συμμετέχουν στο οικονομετρικό υπόδειγμα περιγράφουν υποδομές μεταφορικών δικτύων, γεγονός το οποίο σκιαγραφεί τη σημασία της κατηγορίας των μεταφορικών υποδομών στη διαμόρφωση της μεταβλητότητας που παρουσιάζει η εξαρτημένη μεταβλητή (επίπεδο ευημερίας). Όπως προκύπτει από τις τιμές των τυποποιημένων συντελεστών, η συνεισφορά αυτή είναι της τάξεως του 60% της θετικής συνεισφοράς που έχουν οι εξαρτημένες μεταβλητές στο υπόδειγμα. Η παραπάνω ανάλυση φαίνεται να επικυρώνει την προηγηθείσα επισκόπηση και να αναδεικνύει με ποσοτικούς όρους τη συνεισφορά των υποδομών των μεταφορικών δικτύων στην οικονομική και περιφερειακή ανάπτυξη της χώρας.

**5. Συμπεράσματα**
Η σημασία που κατέχουν οι μεταφορές στην οικονομική ζωή της Ελλάδας είναι δεδομένη και εμφανής σε πολλές εκφάνσεις της κρατικής της υπόστασης και δραστηριότητας, όπως είναι η διοικητική της διάρθρωση, η κατανομή των πόρων, και η παραγωγική της συγκρότηση. Υπό το πρίσμα αυτό, το παρόν άρθρο προσπάθησε να αναδείξει αυτή την σημαντικότητα, εξετάζοντας διαχρονικά και διατομεακά στατιστικά στοιχεία που περιγράφουν ορισμένα θεμελιώδη μακροοικονομικά μεγέθη και μέτρα. Στο στόχαστρο της μελέτης τοποθετήθηκαν οι μεταφορικές υποδομές και τα δίκτυα μεταφορών, τα οποία συνιστούν ένα πάγιο δομημένο κεφάλαιο που εκτείνεται, με διαφορετικές μορφές, στο σύνολο της χώρας. Τα δίκτυα που μελετήθηκαν είναι το οδικό, το σιδηροδρομικό, το ακτοπλοϊκό και το αεροπορικό δίκτυο της Ελλάδας, εστιάζοντας τόσο στη γεωμετρία και τα τεχνικά χαρακτηριστικά τους, όσο και στο ιστορικό, κυκλοφοριακό και πολιτικό τους πλαίσιο.

  Στο μέρος της εμπειρικής ανάλυσης κατασκευάστηκε ένα οικονομετρικό υπόδειγμα πολυμεταβλητής γραμμικής παλινδρόμησης που περιέγραψε το επίπεδο ευημερίας των νομών ως συνάρτηση ενός συνόλου ανεξάρτητων μεταβλητών που σχετίζονται με τις μεταφορικές υποδομές και το κοινωνικοοικονομικό πλαίσιο των νομών της χώρας. Από την ανάλυση προέκυψε η σημαντικότητα των μεταβλητών των μεταφορικών υποδομών στο υπόδειγμα, τόσο ως προς το πλήθος τους όσο και ως προς το μέγεθός τους, η οποία υποδηλώνει τη σημασία της κατηγορίας των μεταφορικών υποδομών στη διαμόρφωση της μεταβλητότητας (όπως περιγράφεται από το υπόδειγμα) που παρατηρείται στο επίπεδο ευημερίας των νομών. Η σημασία αυτή μετράται στο υπόδειγμα ως θετική, εκφράζοντας ότι οι περιφέρειες με υψηλότερο επίπεδο ευημερίας τείνουν γενικά να παρουσιάζουν βελτιωμένη εικόνα στις οδικές και τις λιμενικές τους μεταφορικές υποδομές.

  Ωστόσο, κάτω από μία συνδυασμένη ανάγνωση των προσεγγίσεων της παρούσας μελέτης αναγνωρίζεται ότι ο τρόπος που τα μεταφορικά δίκτυα συντελούν στην προώθηση της περιφερειακής ανάπτυξης διέπεται από υψηλό βαθμό πολυπλοκότητας και πραγματοποιείται σε συνδυασμό με τους υπόλοιπους αναπτυξιακούς παράγοντες μιας περιοχής (γεωγραφική θέση, οικονομίες συγκέντρωσης, τομεακή σύνθεση της τοπικής παραγωγής και απασχόλησης). Όπως προέκυψε από την ανάλυση που προηγήθηκε, η παραγωγική βάση των περιφερειών της χώρας εμφανίζεται να είναι δομημένη κατάλληλα ώστε να επωφελείται θετικά από τις οδικές και λιμενικές υποδομές. Η χαρακτηριστική απουσία των αεροπορικών υποδομών στο υπόδειγμα σκιαγραφεί την ύπαρξη ενός αναπτυξιακού μοντέλου θεμελιωμένου στις βασικές υποδομές.

  Απώτερο σκοπό του άρθρου αποτέλεσε η ανάδειξη, μακροσκοπικά, της δομικής και λειτουργικής διάστασης που συνθέτουν την έννοια των δικτύων μεταφορών, της αναγκαιότητας από κοινού θεώρησής τους στην επιστημονική έρευνα, καθώς και τις διάφορες πτυχές που λαμβάνει η μελέτη των μεταφορικών υποδομών και δικτύων.

## 5. Βιβλιογραφία
*5.1. Ελληνόγλωσση*

# 5. Παράρτημα

## Πίνακας Α
Αρίθμηση των εθνικών οδών της χώρας με την απόφαση του Υπουργού Δημοσίων Έργων με αριθμό Γ.25871/1963

| Αρίθμηση | Διαδρομή |
|---|---|
| 1 | Αθήνα - Δεκέλεια - Αταλάντη - Καμένα Βούρλα - Θερμοπύλες - Λαμία - Στυλίδα - Αλμυρός - Βελεστίνο - Λάρισα - Τέμπη - Κατερίνη - Αλεξάνδρεια - Ν. Χαλκηδόνα - Γέφυρα - Πολύκαστρο - Εύζωνοι. |
| 2 | Κρυσταλλοπηγή (σύνορα με Αλβανία) - Βατοχώρι - Πισοδέρι - Φλώρινα - Έδεσσα - Γιαννιτσά - Νέα Χαλκηδόνα - Θεσσαλονίκη - Λαγκαδίκια - Αμφίπολη - Καβάλα - Τοξότες - Ξάνθη - Πόρτο Λάγος - Κομοτηνή - Μέση - Αλεξανδρούπολη - Φέρρες - Αρδάνιο - Γέφυρα Έβρου. |
| 3 | Ελευσίνα - Θήβα - Λιβαδειά - Μπράλλος - Λαμία - Φάρσαλα - Λάρισα - Τύρναβος - Ελασσόνα - Σέρβια - Κοζάνη - Πτολεμαΐδα - Βεύη - Φλώρινα - Νίκη (σύνορα). |
| 4 | Αλεξάνδρεια - Βέροια - Καστανιά - Πολύμυλος - Κοζάνη. |
| 5 | Ρίο - Αντίρριο - Μεσολόγγι - Αγρίνιο - Αμφιλοχία - Άρτα - Φιλιππιάδα - Ιωάννινα. |
| 6 | Βόλος - Λάρισα - Τρίκαλα - Καλαμπάκα - Γέφυρα Μουργκάνι - Κατάρα - Μέτσοβο - Ιωάννινα - Ηγουμενίτσα. |
| 7 | Κόρινθος - Νεμέα - Άργος - Τρίπολη - Μεγαλόπολη - Καλαμάτα. |
| 8 | Αθήνα - Κόρινθος - Ξυλόκαστρο - Δερβένι - Αίγιο - Ρίο - Πάτρα |
| 9 | Πάτρα - Κάτω Αχαΐα - Λεχαινά - Πύργος - Ζαχάρω - Κυπαρισσία - Πύλος - Μεθώνη |
| 12 | Θεσσαλονίκη - Σέρρες - Μεσορράχη - Δράμα - Καβάλα |
| 13 | Κατερίνη - Άγιος Δημήτριος - Ελασσόνα |
| 14 | Δράμα - Παρανέστι - Σταυρούπολη - Ξάνθη. |
| 15 | Γέφυρα Μουργκάνι - Γρεβενά - Μπάρα - Νεάπολη - Καστοριά - Τρίγωνο - Άγιος Γερμανός |
| 16 | Θεσσαλονίκη - Αρναία - Ιερισσός |
| 17 | Ιωάννινα - Δωδώνη |
| 18 | Εθνική Οδός 21 - Καναλάκι - Παραμυθιά - Μενίνα. |
| 20 | Κοζάνη - Μπάρα - Νεάπολη - Τσοτύλι - Πεντάλοφος - Επταχώρι - Κόνιτσα - Καλπάκι - Ιωάννινα |
| 21 | Φιλιππιάδα - Πρέβεζα. |
| 22 | Καλπάκι - Κακαβιά (Αλβανία). |
| 24 | Κέρκυρα - Παλαιοκαστρίτσα. |
| 25 | Κέρκυρα - Γύρος Αχιλλείου. |
| 26 | Ελασσόνα - Δεσκάτη - Καρπερό. |
| 27 | Άμφισσα - Μπράλλος. |
| 28 | Οδός Αεροδρομίου Λάρισας |
| 29 | Στενή - Όσιος Λουκάς |
| 30 | Άρτα - Βουλγαρέλι - Τρίκαλα - Καρδίτσα - Νέο Μοναστήρι - Φάρσαλα - Μικροθήβες - Αγχίαλος - Βόλος |
| 31 | Αίγιο - Φτέρη - Καλάβρυτα |
| 33 | Πάτρα - Τριπόταμο - Λεβίδι |
| 34 | Βόλος - Νεοχώρι - Τσαγκαράδα - Χορευτό |
| 35 | Ζάκυνθος - Κερί. |
| 36 | Μυτιλήνη - Καλλονή |
| 38 | Λαμία - Καρπενήσι - Αγρίνιο - Θέρμο |
| 39 | Τρίπολη - Σπάρτη - Γύθειο |
| 40 | Οδός Αεροδρομίου Αγρινίου |
| 42 | Αμφιλοχία - Βόνιτσα - Λευκάδα |
| 44 | Θήβα - Χαλκίδα - Αλιβέρι - Λέπουρα |
| 46 | Οδός Αεροδρομίου Τανάγρας |
| 48 | Λιβαδειά - Αράχοβα - Δελφοί - Άμφισσα - Λιδωρίκι - Ναύπακτος - Αντίρριο. |
| 50 | Αργοστόλι - Σάμη. |
| 51 | Αρδάνιο - Διδυμότειχο - Ορεστιάδα - Καστανιές - Τουρκικά σύνορα. |
| 53 | Αλεξανδρούπολη - Αισύμη - Δέρειο - Βουλγαρικά σύνορα |
| 54 | Αθήνα - Σταυρός - Ραφήνα |
| 55 | Ξάνθη - Εχίνος - Βουλγαρικά σύνορα |
| 56 | Αθήνα - Πειραιάς |
| 57 | Δράμα - Κάτω Νευροκόπι - Βουλγαρικά σύνορα. |
| 58 | Οδός Αεροδρομίου Ελευσίνας. |
| 59 | Μεσορράχη - Αμφίπολη. |
| 60 | Οδός Αεροδρομίου Μεγάρων |
| 62 | Σάμος - Λιμένας Καρλοβασίου. |
| 63 | Σέρρες - Σιδηρόκαστρο - Προμαχώνας. |
| 64 | Κάτω Αχαΐα - Άραξος |
| 65 | Εθνική Οδός 2 (στη Θεσσαλονίκη) - Κιλκίς |
| 66 | Εθνική Οδός 7 (στο σιδηροδρομικό σταθμό Νεμέας) - Νεμέα - Ψάρι - Σκοτεινή - Κανδήλα - Λεβίδι. |
| 67 | Οδός Αερολιμένα Μακεδονίας (Θεσσαλονίκη). |
| 68 | Φίχτια - Μυκήνες. |
| 69 | Λιμένας - Λιμενάρια Θάσου. |
| 70 | Άργος - Ναύπλιο - Θέατρο Επιδαύρου - Παλαιά Επίδαυρος |
| 71 | Οδός Αεροδρομίου - Νέας Αγχιάλου |
| 72 | Οδός Αεροδρομίου Τριπόλεως |
| 74 | Τρίπολη - Λεβίδι - Βυτίνα - Ολυμπία - Βαρβάσαινα - Πύργος |
| 75 | Καλλιμασιά - Χίος - Καρδάμυλα |

| Αρίθμηση | Διαδρομή |
|---|---|
| 76 | Μεγαλόπολη - Ανδρίτσαινα - Ναός Επικουρείου Απόλλωνος |
| 77 | Χαλκίδα - Ιστιαία - Αιδηψός |
| 78 | Οδός Αεροδρομίου Ανδραβίδας. |
| 79 | Άγιος Μερκούριος - Σκάλα Ωρωπού |
| 80 | Οδός Αεροδρομίου Άρεως |
| 81 | Άνοιξη - Καπανδρίτι - Κάλαμος - Αμφιαράεον. |
| 82 | Σπάρτη - Καλαμάτα - Μεσσήνη - Βελίκα - Χατζή - Πύλος |
| 83 | Αθήνα - Κηφισιά - Άνοιξη - Μαραθώνας - Ραφήνα |
| 84 | Σπάρτη - Μυστράς |
| 85 | Ραφήνα - Πόρτο Ράφτη - Λαύριο |
| 86 | Κροκεές - Μολάοι - Μονεμβασιά. |
| 87 | Γλυκά Νερά - Παλλήνη - Χριστούπολη - Σπάτα |
| 88 | Οδός Αεροδρομίου Μαριτσών |
| 89 | Σταυρός - Παιανία - Μαρκόπουλο - Λαύριο - Προέκταση προς Κ. Ποσειδωνία - Σούνιο |
| 90 | Καστέλλι - Χανιά - Ρέθυμνο - Ηράκλειο - Άγιος Νικόλαος - Σητεία |
| 91 | Αθήνα - Λεωφόρος Συγγρού - Γλυφάδα - Βουλιαγμένη - Βάρκιζα - Λαγονήσι - Σαρωνίδα - Παλ. Φώκαια - Σούνιο - Προέκταση προς Κ. Ποσειδωνία - Λαύριο |
| 92 | Οδός Αεροδρομίου Καστελλίου - Ανάληψη Χερσονήσσου Ηρακλείου |
| 94 | Οδός Αεροδρομίου Σούδας - Χανίων |
| 95 | Ρόδος - Κολύμπια - Λίνδος |
| 97 | Ηράκλειο - Φαιστός - Αγία Γαλήνη |
| 99 | Ηράκλειο - Αρκαλοχώρι |

(πηγή: Απόφαση Υπουργού Δημοσίων Έργων με αριθμό Γ.25871/1963)

**Πίνακας Β**
Οι βασικοί οδικοί άξονες στην Ελλάδα

| α/α | Ονομασία | Διαδρομή | Κωδικός | Μήκος (km) | Κατασκευή (km) |
|---|---|---|---|---|---|
| 1 | ΠΑΘΕ | Εύζωνοι – Θεσσαλονίκη – Λάρισα – Λαμία – Αθήνα – Πειραιάς – Ελευσίνα – Μέγαρα – Κόρινθος – Αίγιο – Ρίο | Α1 + Α8 | 770 | 521 |
| 2 | Εγνατία Οδός | Ηγουμενίτσα – Ιωάννινα – Γρεβενά – Κοζάνη – Βέροια – Θεσσαλονίκη – Καβάλα – Ξάνθη – Κομοτηνή – Αλεξανδρούπολη – Κήποι Έβρου | Α2 | 670 | 546 |
| 3 | Ιόνια Οδός | Ιωάννινα – Άρτα – Αμφιλοχία – Αγρίνιο – Μεσολόγγι – Γέφυρα Ρίου-Αντιρρίου – Ρίο- Πάτρα – Πύργος Ηλείας –Τσακώνα Μεσσηνίας | Α5 | 383 | 19 |
| 4 | Ε65 | Παναγιά Τρικάλων – Καλαμπάκα – Τρίκαλα – Καρδίτσα – Λαμία | Α3 | 175 | 0 |
|  | Ε65 | Λαμία – Αντίρριο |  | 172 | 0 |
| 5 | Κεντρικός Άξονας Πελοποννήσου | Κόρινθος – Τρίπολη – Μεγαλόπολη – Καλαμάτα | Α7 | 201 | 62 |
| 6 | Βόρειος Άξονας Κρήτης | Κίσσαμος – Χανιά – Ρέθυμνο – Ηράκλειο – Άγιος Νικόλαος – Σητεία | Α90 | 296 | 195 |
| Σύνολο |  |  |  | 2485 | 1343 |

(πηγές: ΥΠΕΘΟ, 1993; Εγνατία Οδός ΑΕ, 2008; ΥΠΕΧΩΔΕ, 2006, 2008)

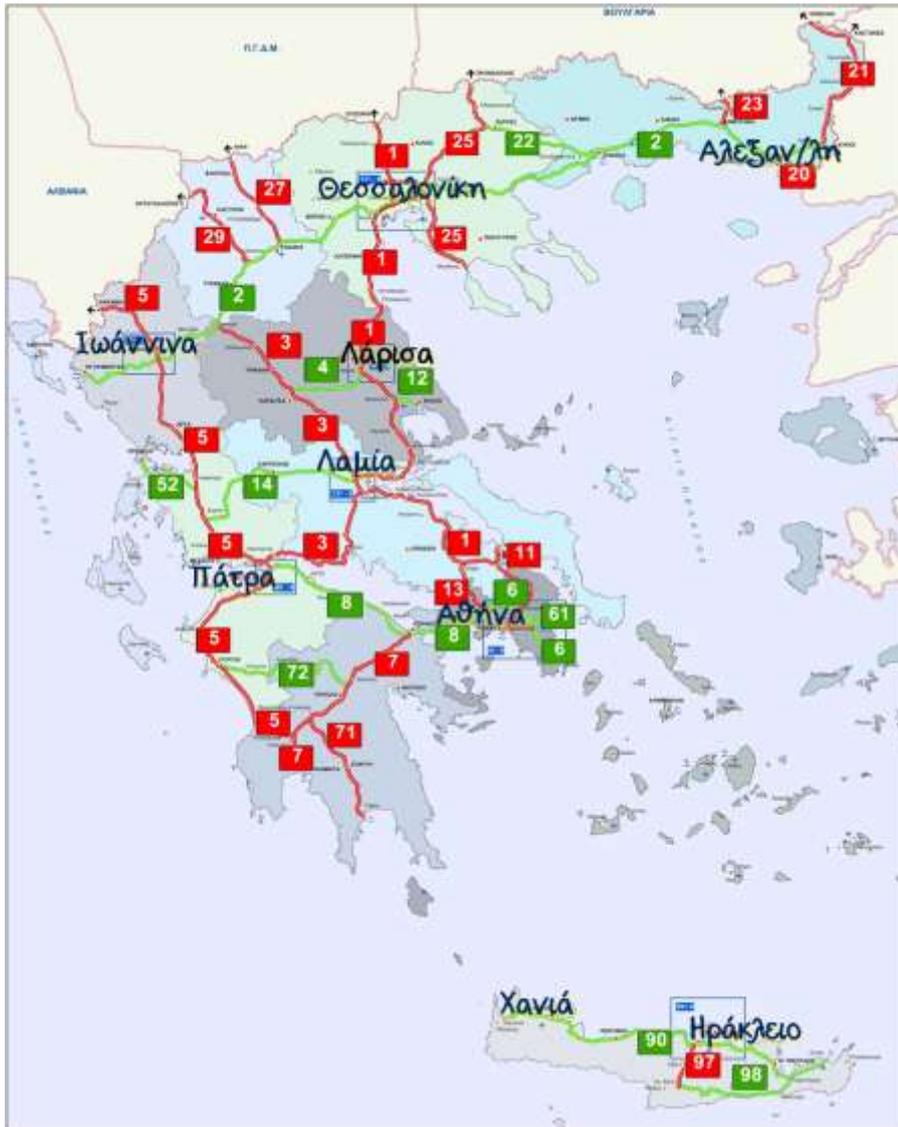

**Σχήμα (α).** Οι βασικοί οδικοί άξονες στην Ελλάδα (πηγή: ΥΠΕΧΩΔΕ, 2008, ίδια επεξεργασία).

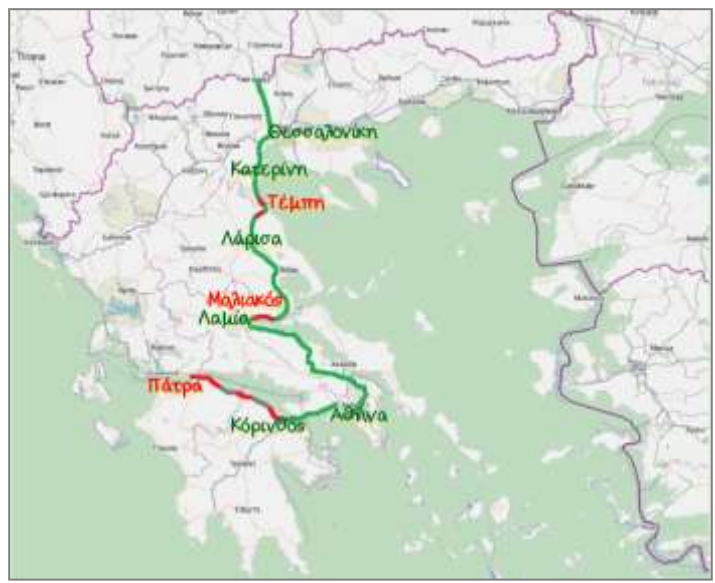

**Σχήμα (β).** Ο αυτοκινητόδρομος Αθήνας – Θεσσαλονίκης. Η κατασκευή των τμημάτων του Μαλιακού και των Τεμπών βρίσκεται σε εξέλιξη (πηγή: http://upload.wikimedia.org/wikipedia/el/b/b7/A1_Map.jpg, προσπελάστηκε 03/04/2015).

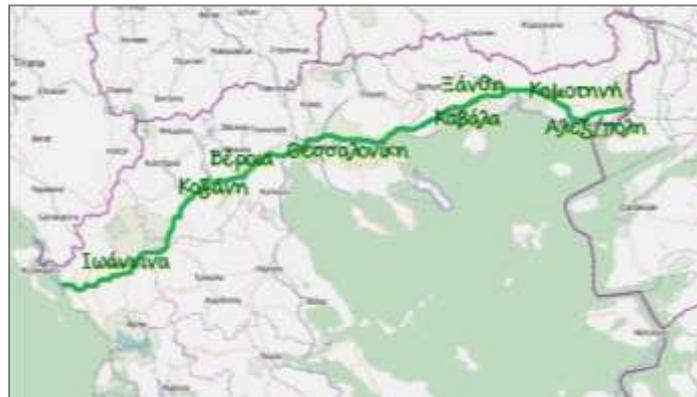

**Σχήμα (γ).** Η Εγνατία Οδός (πηγή: http://upload.wikimedia.org/wikipedia/el/0/05/A2_Map.jpg, προσπελάστηκε 03/04/2015).

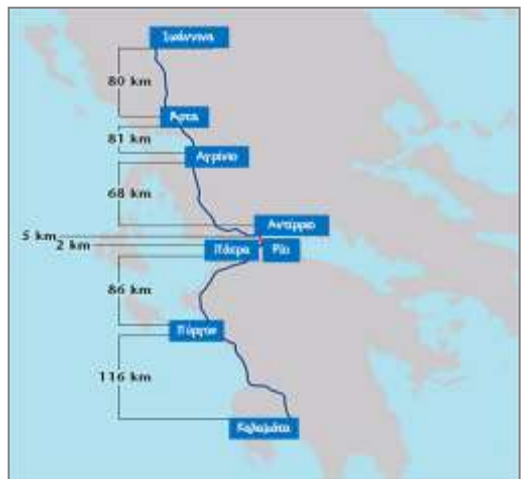

**Σχήμα (δ).** Η Ιόνια Οδός (πηγή: Polyzos et al., 2014).

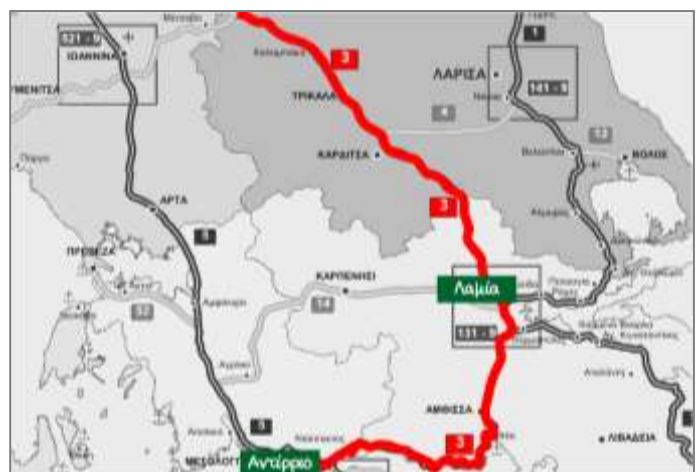

**Σχήμα (ε).** Η Οδός Ε65 (πηγή: ΥΠΕΧΩΔΕ, 2008).

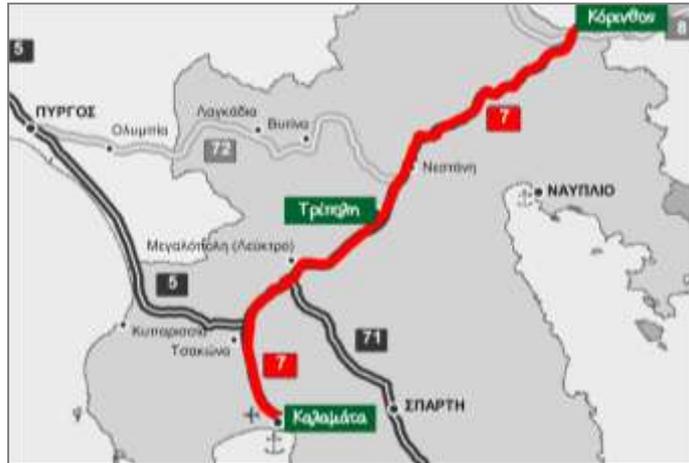

**Σχήμα (στ).** Ο κεντρικός οδικός άξονας Πελοποννήσου (πηγή: ΥΠΕΧΩΔΕ, 2008).

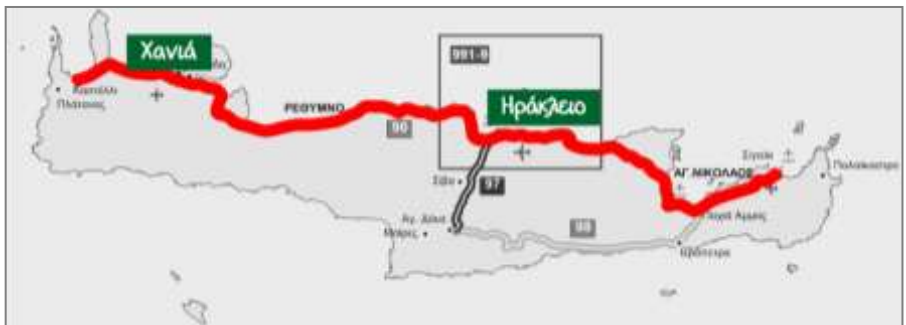

**Σχήμα (ζ).** Ο Βόρειος Άξονας Κρήτης (πηγή: ΥΠΕΧΩΔΕ, 2008).

σελ. | 40